%% file: PRDSergiVubCombinedarXiv.tex
\documentclass[aps,twocolumn,showpacs,preprintnumbers,amsmath,amssymb,nofootinbib,showkeys]{revtex4-1}

\RequirePackage[T1]{fontenc}

\usepackage{epsfig,graphicx,amsmath,amssymb}
\RequirePackage{mathptmx}      
\RequirePackage{flushend}
\RequirePackage{color}
\usepackage{changes}
\usepackage{multirow}
\RequirePackage{hyperref}
\hypersetup{
	linktocpage,
    colorlinks,
    citecolor=blue,
    filecolor=black,
    linkcolor=blue,
    urlcolor=blue,
}

\begin{document}

\title{Pad\'{e} approximants to $B\to\pi\ell\nu_{\ell}$ and $B_{s}\to K\ell\nu_{\ell}$ and determination of $|V_{ub}|$}

\author{Sergi Gonz\`{a}lez-Sol\'{i}s$^{1,2}$}\email{sgonzal@iu.edu}
\author{Pere Masjuan$^{3,4}$} \email{masjuan@ifae.es}
\author{Camilo Rojas$^{3,4}$} \email{crojas@ifae.es}

\affiliation{$^1$Department of Physics, Indiana University, Bloomington, IN 47405, USA\\
$^2$Center for Exploration of Energy and Matter, Indiana University, Bloomington, IN 47408, USA\\
$^3$Grup de F\'{i}sica Te\`{o}rica, Departament de F\'{i}sica, Universitat Aut\`{o}noma de Barcelona\\
$^3$Institut de F\'{i}sica d'Altes Energies (IFAE) and The Barcelona Institute of Science and Technology, Universitat Aut\`{o}noma de Barcelona, E-08193 Bellaterra (Barcelona), Spain}

\begin{abstract}

In light of the first observation of the semileptonic decay $B^{0}_{s}\to K^{-}\mu^{+}\nu_{\mu}$ by the LHCb collaboration, we revisit the determination of the CKM parameter $|V_{ub}|$ from exclusive semileptonic $B$-meson decays.
A controlled theoretical input on the Standard Model $B\to\pi$ and $B_{s}\to K$ vector and scalar form factors from Lattice QCD in the large $q^{2}$ region, in combination with experimental measurements of the differential $B\to\pi\ell\nu_{\ell}$ and $B^{0}_{s}\to K^{-}\mu^{+}\nu_{\mu}$ branching ratio distributions, has allowed us determine $|V_{ub}|=3.86(11)\times10^{-3}$ and $|V_{ub}|=3.58(9)\times10^{-3}$ from the analyses of the individual decay channels, respectively, and $|V_{ub}|=3.68(5)\times10^{-3}$ from a simultaneous analysis of both decays, which is only a $1.4\%$ error and differs by $1.8\sigma$ with respect to the value from inclusive determinations $|V_{ub}|=4.25(12)^{+15}_{-14}(23)\times10^{-3}$.
Our results are based on the use of Pad\'{e} approximants to the participating form factors, highlight the importance of the decay $B_{s}\to K\mu\nu_{\mu}$ in complementing the traditional $B\to\pi\ell\nu_{\ell}$ one in the exclusive determination of $|V_{ub}|$, and allow obtain, to the best of our knowledge, the first correlated results for the $B\to\pi$ and $B_{s}\to K$ vector and scalar form factors.
We hope that our study strengthens the case for precise measurements of the differential $B_{s}\to K\ell\nu_{\ell}$ decay rate with a finer resolution of the $q^{2}$ bins, as it would definitely allow achieving more conclusive results for $|V_{ub}|$.


\end{abstract}


\maketitle

\input{Section1Introduction.tex}
\input{Section2Lagrangian.tex}
\input{Section3Fits.tex}
\input{Section4Pheno.tex}
\input{SectionLastOutlook.tex}

\begin{acknowledgements}

The work of S.G-S. has been supported in part by by the National Science Foundation under the Grant no. PHY-2013184 and the U.S. Department of Energy under the Grant no. DE-FG02-87ER40365. 
P.M and C.R. have received funding from the Spanish Ministry of Science and Innovation (PID2020-112965GB-I00/AEI/ 10.13039/501100011033) and from the
Agency for Management of University and Research Grants of the Government of Catalonia (project
SGR 1069).

\end{acknowledgements}

\appendix

\input{Appendix1.tex}

\input{SectionBibliography.tex}
\end{document}

%% file: Section1Introduction.tex
\section{Introduction}\label{section1}

The Cabibbo-Kobayashi-Maskawa (CKM) matrix describes quark flavor-changing transitions in the Standard Model (SM).
The elements of the CKM matrix, denoted by $V_{ij}$ for a transition of a $j$-type quark to a $i$-type ones, are fundamental parameters of the SM and knowledge of their magnitude with high accuracy is absolutely mandatory for precise SM tests.
The CKM matrix is unitary in the SM, {\it{i.e.}} it satisfies $\sum_{i}V_{ij}V_{ik}^{*}=\delta_{jk}$ and $\sum_{j}V_{ij}V_{kj}^{*}=\delta_{ik}$.
Violations of unitarity are evidence of physics beyond the Standard Model (BSM). 
Each particular matrix element can be determined from multiple processes, and if the SM predictions do not imply identical values of the particular element, that could also be a hint for non-SM physics.
Of course, to  unravel such BSM's evidences require precision calculations of the SM.

There are many processes where to test the CKM matrix and extract its elements. 
Among them, purely leptonic weak decays, {\it{e.g.}} $P^{-}\to\ell^{-}\bar{\nu}_{\ell}$ with $P=\{\pi,K,D,B\}$, offer (in general) a theoretically clean environment for the determination of the CKM elements more advantageous than the semileptonic ones,\footnote{The only hadronic input required in leptonic decays are the decay constants of the decaying mesons, which are well calculated in Lattice QCD \cite{Aoki:2019cca}.} 
where the decay rates depend on hadronic information that is encoded in form factors. 
In addition, both leptonic and semileptonic decays offer an opportunity to test lepton flavor universality as $\ell$ can be $e,\mu$ or $\tau$.
The current status of the magnitude of the CKM matrix elements and future prospects for improving their determination can be found in the Particle Data Group \cite{Zyla:2020zbs} as well in the Flavour Lattice Averaging Group (FLAG) report \cite{Aoki:2019cca} (see also Ref.\,\cite{Gottlieb:2020zsa}).

In this paper we concentrate on $|V_{ub}|$, one of the least-known CKM elements which governs the strength of $b\to u$ transitions, and we are going to consider only exclusive processes.
Among the three possible $B$-meson leptonic channels to obtain exclusive determinations of $|V_{ub}|$, the only available experimental input comes from $B\to\tau\nu_{\tau}$, since the partial decay rates to $e$ and $\mu$ have not been measured yet.
However, the averaged experimental measurements \cite{Aoki:2019cca} from BaBar, $BR(B\to\tau\nu_{\tau})=1.79(48)\times10^{-4}$, and Belle, $BR(B\to\tau\nu_{\tau})=0.91(22)\times10^{-4}$, both coming from averaging different $\tau$-reconstruction channels, do not agree well and have large errors (about $25\%$).
These measurements yield $|V_{ub}|f_{B}=0.72(9)$ MeV and $|V_{ub}|f_{B}=1.01(14)$ MeV \cite{Aoki:2019cca}, respectively, which can be used to extract $|V_{ub}|$ when combined with Lattice-QCD predictions of the $B$-meson decay constant $f_{B}$.  
As an example, using $f_{B}=192.0(4.3)$ MeV from a $N_{f}=2+1$ flavor gauge-field ensemble \cite{Aoki:2019cca}, one gets $|V_{ub}|=5.26(12)(73)\times10^{-3}$, from the BaBar measurement, and $|V_{ub}|=3.75(8)(47)\times10^{-3}$, from the Belle one, where the first uncertainty comes from the error in $f_{B}$ and the second one from experimental considerations.
The discrepancy between these two results is manifest.
This means, in practice, that a reliable determination of $|V_{ub}|$ from leptonic decays will only be possible with the new and more precise data expected from Belle-II \cite{Kou:2018nap}.

Currently, the most precise determination of $|V_{ub}|$ comes from charmless semileptonic $B$-meson decays, using exclusive or inclusive methods.
Inclusive determinations rely on the operator product expansion and perturbative QCD applied to $B\to X_{u}\ell\bar{\nu}_{\ell}$ observables, while the exclusive one require knowledge of the participating form factors.
The most competitive exclusive determination of $|V_{ub}|$ is obtained from the decay channel $B\to\pi\ell\nu_{\ell}$, which has generally exhibited a tension with inclusive determinations (see \cite{Gottlieb:2020zsa} for a history of the comparison).
More specifically, the experimental $B\to\pi\ell\nu_{\ell}$ observable depends upon know quantities, $|V_{ub}|$ -that we would like to determine- and the $B\to\pi$ form factors, that we need to describe and extrapolate to $q^2=0$ to obtain that $|V_{ub}|$.
While QCD light-cone sum rules have been used to calculate the value of the vector form factor at $q^{2}=0$ with certain error \cite{Bharucha:2012wy}, precise Lattice-QCD simulations are available in the energy region close to the maximum momentum transfer to the leptons, $17$  GeV$^2$$<q^{2}<26$ GeV$^{2}$, from the HPQC Collaboration \cite{Dalgic:2006dt}, the RBC and UKQCD (RBC/UKQCD) Collaborations \cite{Flynn:2015mha}, and the Fermilab Lattice and MILC (FNAL/MILC) Collaborations \cite{Lattice:2015tia}.
Several representations have been proposed for the form factor interpolation between these two regimes, including dipole-like functions \cite{Becirevic:1999kt,Ball:2004ye}, the so called $z$-expansion parameterizations \cite{Boyd:1994tt,Bourrely:2008za}, and more recently Pad\'{e} approximants \cite{Gonzalez-Solis:2018ooo}.
These parameterizations can be used to obtain $|V_{ub}|$ via a simultaneous fit of the Lattice-QCD form factor calculations and the partial branching ratios experimental data \cite{Adam:2007pv,delAmoSanchez:2010af,Lees:2012vv,Ha:2010rf,Sibidanov:2013rkk}.
The $q^{2}$ dependence of the form factor is thus fixed at small $q^{2}$ by data, which due to phase-space suppression have poor access to the large-$q^{2}$ region, and at large $q^{2}$ by the Lattice simulations, which has a larger uncertainty than experiment at small $q^{2}$ due to the extrapolation.
The theoretical uncertainties on the form factors were the dominant source error in $|V_{ub}|$ until the 2015 FNAL/MILC results \cite{Lattice:2015tia}, which brought the QCD error to the same level as the experimental one.
In the intermediate energy region around $q^{2}\sim20$ GeV$^{2}$, both the experimental and Lattice-QCD errors are similar in size.
This region is decisive for determining $|V_{ub}|$ with precision, and can be employed to estimate the individual contributions from experimental and Lattice data.

The semileptonic $B_{s}\to K\ell\nu_{\ell}$ also depends on the CKM element $|V_{ub}|$.
The only difference with respect to the decay $B\to\pi\ell\nu_{\ell}$ is that in $B_{s}\to K\ell\nu_{\ell}$ the light spectator quark is a strange quark instead of an up or down quark as in the former process.
The $B_{s}\to K$ form factors have been simulated on the Lattice by the HPQCD Collaboration \cite{Bouchard:2014ypa}, the RBC/UKQCD Collaborations \cite{Flynn:2015mha}, the ALPHA Collaboration \cite{Bahr:2016ayy}, and more recently by the FNAL/MILC Collaborations \cite{Bazavov:2019aom}.
As in the $B\to\pi\ell\nu_{\ell}$ case, these calculations can be used to extract $|V_{ub}|$ when combined with experimental measurements for $B_{s}\to K\ell\nu_{\ell}$, which can play an important role in reassessing the result and addressing the current exclusive versus inclusive $|V_{ub}|$ puzzle.
Recently, the first experimental data on $B_{s}\to K\ell\nu_{\ell}$ became available by the LHCb Collaboration, which measured the partial branching ratio distribution in two regions of $q^{2}$ \cite{Aaij:2020nvo}.
In our work, we use these data to determine $|V_{ub}|$ and illustrate the potential of a combined analysis of the decays $B\to\pi\ell\nu_{\ell}$ and $B_{s}\to K\ell\nu_{\ell}$. 
The decay $B_{s}\to K\ell\nu_{\ell}$ is also expected to be studied at the Belle-II experiment \cite{Kou:2018nap}, where the $e^{+}e^{-}$ collisions would yield a cleaner environment than the LHC.  
Other processes offering interesting information on $|V_{ub}|$, but not considered in our analysis, include the $B_{\ell4}$ \cite{Kang:2013jaa} and the baryonic $\Lambda_{b}\to p\ell\bar{\nu}_{\ell}$ decays \cite{Aaij:2015bfa,Detmold:2015aaa}.

This paper is structured as follows.
The hadronic matrix element and the participating vector and scalar form factors are defined in Sec. \ref{section2}, where the differential decay distribution in terms of the latter is also given.
In Sec. \ref{section3}, we determine $|V_{ub}|$ and the corresponding form factor parameters from fits to the $B\to\pi\ell\nu_{\ell}$ and $B_{s}\to K\mu\nu_{\mu}$ experimental measurements on the differential branching ratio distribution combined with the Lattice-QCD theoretical information on the form factors.  
In Secs.\,\ref{section3sub1} and \ref{section3sub2}, we first perform individual studies of both decays separately, and after that, in Sec.\,\ref{section3sub3} we perform a simultaneous analysis including all available experimental and theoretical information on both exclusive decays.
The outputs of our fits are then used in Sec.\,\ref{section4} to calculate some interesting phenomenological observables such as total decay rates, $\tau$-to-$\mu$ ratio of differential decay rates and the forward-backward asymmetry.
We close with an outlook in Sec. \ref{conclusions}.

%% file: Section2Lagrangian.tex
\section{Decay amplitude and form factors}\label{section2}

In the SM, the amplitude for the exclusive semileptonic decays $B\to\pi\ell\nu_{\ell}$ is given by:
\begin{eqnarray}
i\mathcal{M}=\frac{G_{F}V_{ub}}{\sqrt{2}}L_{\mu}H^{\mu}\,,
\label{Eq:amplitude}
\end{eqnarray}
where $G_{F}$ is the Fermi constant and $V_{ub}$ is the participating element of the CKM matrix.
In Eq.\,(\ref{Eq:amplitude}), the leptonic current have the structure
\begin{eqnarray}
L_\mu&=&\bar{u}(p_{\nu})\gamma_\mu (1-\gamma^5)v(p_{\ell})\,,
\end{eqnarray}
while the hadronic matrix element can be decomposed in terms of allowed Lorentz structures and two form factors encoding the hadronic information:
\begin{eqnarray}
H_{\mu}&=&\langle\pi(p_{\pi})|\bar{u}\gamma_{\mu}b|B({p_{B}})\rangle\nonumber\\[1ex]
&=&\left(p_{B}+p_{\pi}-q\frac{m_{B}^{2}-m_{\pi}^{2}}{q^{2}}\right)_{\mu}f_{+}(q^{2})+\frac{m_{B}^{2}-m_{\pi}^{2}}{q^{2}}q_{\mu}f_{0}(q^{2})\,,\nonumber\label{Eq:VectorCurrent}\\[1ex]
\end{eqnarray}
where $q_\mu = (p_{B}-p_{\pi})_\mu =(p_{\ell}+p_{\nu_{\ell}})_\mu$ is the transferred momentum to the dilepton pair.
The $q^{2}$ functions $f_{+}(q^{2})$ and $f_{0}(q^{2})$ are, respectively, the vector and scalar form factors corresponding to the exchange of $J^{P}=1^{-}$ and $0^{+}$ particles in case there is non-resonant background.       
These two form factors satisfy a kinematical constraint,
\begin{equation}
f_{+}(0)=f_{0}(0)\,,
\label{Eq:FormFactorConstraint}
\end{equation}
which eliminates the (spurious) pole at $q^{2}=0$ in Eq.\,(\ref{Eq:VectorCurrent}).

In terms of these form factors, the dilepton mass squared distribution reads:
\begin{widetext}
\begin{eqnarray}
\frac{d\Gamma(B\to\pi\ell\nu_{\ell})}{dq^{2}}&=&\frac{G_{F}^{2}|V_{ub}|^{2}\lambda^{1/2}(m_{B}^{2},m_{\pi}^{2},q^{2})}{128m_{B}^{3}\pi^{3}q^{2}}\left(1-\frac{m_{\ell}^{2}}{q^{2}}\right)^{2}\times\left\{m_{\ell}^{2}(m_{B}^{2}-m_{\pi}^{2})^{2}|f_{0}(q^{2})|^{2}+\frac{2q^{2}}{3}\lambda(m_{B}^{2},m_{\pi}^{2},q^{2})\left(1+\frac{m^{2}_{\ell}}{2q^{2}}\right)|f_{+}(q^{2})|^{2}\right\}\,,\nonumber\\[1ex]
\label{Eq:DecayDistribution}
\end{eqnarray}
\end{widetext}
where $\lambda(x,y,z)=(x+y-z)^{2}-4xy$ is the Kallen function.
For the decay $B_{s}\to K\ell\nu_{\ell}$, the distribution is that of Eq.\,(\ref{Eq:DecayDistribution}) but replacing $m_{B}\to m_{B_{s}}, m_{\pi}\to m_{K}$ and the $B\to\pi$ form factors by the $B_{s}\to K$ ones.

The present best knowledge of the vector and scalar $B\to\pi$ and $B_{s}\to K$ form factors are obtained from Lattice-QCD calculations in the large-$q^{2}$ region,
which are then extrapolated to the full kinematic range, {\it{i.e.}} $0<q^{2}<(m_{B}-m_{\pi})^{2}$,
using parametrizations based on resonance-exchange ideas \cite{Wirbel:1985ji,Grinstein:1986ad,Nussinov:1986hw,Suzuki:1987ap} or the $z$-expansion \cite{Bourrely:2008za}.
As shown in \cite{Gonzalez-Solis:2018ooo}, these parametrizations are in a form or another a certain kind of Pad\'{e} approximant, which we will use in this work.
Here, we only briefly review them, referring to Refs.\,\cite{Baker,Gonzalez-Solis:2018ooo} for further details. 

Pad\'{e} approximants (PA in what follows) to a given function are ratios of two polynomials (with degree $M$ and $N$, respectively)
\begin{eqnarray}
\hspace{-0.5cm}P^{M}_{N}(q^{2})=\frac{\sum_{j=0}^{M}a_{j}(q^{2})^{j}}{\sum_{k=0}^{N}b_{k}(q^{2})^{k}}=
\frac{a_{0}+a_{1}q^{2}+\cdots+a_{M}(q^{2})^{M}}{1+b_{1}q^{2}+\cdots+b_{N}(q^{2})^{N}}\,,
\label{Eq:Pade}
\end{eqnarray}
with coefficients determined after imposing a set of a accuracy-through-order conditions with the function $f(q^{2})$ one wants to approximate: 
\begin{equation}
f(q^2)-P^{M}_{N}(q^{2})={\mathcal O}(q^2)^{M+N+1}\,.
\label{Eq:order}
\end{equation}

In our case, the key point is to realize that the form factors $f_{+,0}(q^2)$ are Stieltjes functions, which are functions that can be represented by an integral form defined as~\cite{Baker}
\begin{equation}\label{Stieltjes}
f(q^2) = \int_0^{1/R} \frac{d \phi(u)}{1-u q^2}\,,
\end{equation}
where $\phi(u)$ is any bounded and non-decreasing function. 
By defining $R=s_{\rm th}=(m_B+m_{\pi})^2$, or $(m_{B_s}+m_{K})^2$ for $B_s \to K \ell \nu_{\ell}$, identifying 
$d\phi(u) = \frac{1}{\pi}\frac{{\rm{Im}}f(1/u)}{u}du$, and making the change of variables $u=1/s$, Eq.~(\ref{Stieltjes}) returns a dispersive form factor representation 
\begin{equation}
f(q^{2})=\frac{1}{\pi}\int_{s_{\rm{th}}}^{\infty}ds^{\prime}\frac{{\rm{Im}}f(s^{\prime})}{s^{\prime}-q^{2}-i\varepsilon}\,,
\label{dispersionrelation}
\end{equation}
where $q^{2}$ is the invariant mass of the lepton pair. 
Since $f(q^{2})$, and its imaginary part, is created by the vector current, Im$f(s)$ is a positive function (Im$f(s) = \pi \rho(s)$, and $\rho(s)$ the spectral function), the requirement of $\phi(u)$ to be non-decreasing is fulfilled and the convergence of PA to $f(q^{2})$ is guaranteed.

Whenever information on resonance contributions to those form factors is available, for example the position of the resonance in the complex $q^2$ plane, it can be easily included in the definition of the PA by forcing the poles of the approximant to lie exactly at the position of the resonance. 
When the $N$ poles are included in advance, the PA is called Pad\'{e}-Type $T^M_N$, while when $K<N$  poles are fixed and the rest $N-K$ are left free, it is called Partial-Pad\'{e} approximant, $P^M_{K,N-K}$. 
In the present case where $B^*(1^-)$ resonance is known and can be nicely parametrized with the narrow-width approximation (the resonances lies in the real axis), we will also consider such PA extensions.

In the present work we are going to use Pad\'{e} theory extensively to parametrize both $B \to \pi$ and $B_s \to K$ vector and scalar form-factors in order to extrapolate the large-$q^2$ region's calculations obtained from Lattice-QCD to the full kinematic range and, in particular, at $q^2=0$. An advantage of the Pad\'{e} method in front of other parameterizations is the monitoring of unitary violations. While the unitary constraint in $z$-parameterizations is rather vague, with PA it is crystal clear \cite{Gonzalez-Solis:2018ooo,Masjuan:2008cp,Masjuan:2009wy}: PA to Stieltjes functions are also Stieltjes functions. 
All PA poles must be real. 
The presence of complex-conjugated poles and/or zeros when approximating Stieltjes functions is a notorious violation of convergence, possible only if unitary violation is present in data (which is a non-Stieltjes property). 
We will explore this property in the present work which extends thus supersedes our previous attempt in Ref. \cite{Gonzalez-Solis:2018ooo}.



%% file: Section3Fits.tex
\section{$|V_{ub}|$ determinations}\label{section3}

\subsection{Fits to the decay $B\to\pi\ell\nu_{\ell}$}\label{section3sub1}

We start performing fits to the $B\to\pi\ell\nu_{\ell}$ differential branching ratio distribution experimental measurements combined with the $B\to\pi$ form factor Lattice-QCD simulated data.
To that end, we minimize the following $\chi^{2}$-like function,
\begin{eqnarray}
\chi^{2}_{B\pi}&=&N\left(\frac{\chi^{2}_{\rm{data}}}{N_{\rm{data}}}+\frac{\chi^{2}_{\rm{Lattice}}}{N_{\rm{Lattice}}}\right)\,,
\label{Eq:chi2}
\end{eqnarray}
where $N_{\rm{data}}$ is the number of experimental points, $N_{\rm{Lattice}}$ the number of the Lattice form factor $q^{2}$-points, and $N=N_{\rm{data}}+N_{\rm{Lattice}}$.
The above definition ensures the $\chi^{2}$ function with a smaller number of points is well represented in $\chi^{2}_{B\pi}$, and is not overridden by that with a larger number of points.
The individual $\chi^{2}$ functions in Eq.\,(\ref{Eq:chi2}) are given by:
\begin{equation}
\chi^{2}_{\rm{data}}=\sum_{i,j=1}^{13}\Delta_{i}^{{\rm{data}}}({\rm{Cov}}_{ij}^{\rm{data}})^{-1}\Delta_{j}^{\rm{data}}\,,
\label{Eq:chi2data}
\end{equation}
where
\begin{equation}
\Delta_{k}^{\rm{data}}=\left(\frac{\Delta B}{\Delta q^{2}}\right)_{k}^{\rm{data}}-\frac{\tau_{B^{0}}}{\Delta q_{k}^{2}}\int_{q_{k}^{\rm{low}}}^{q_{k}^{\rm{high}}}dq^{2}\frac{d\Gamma}{dq^{2}}\,,
\label{Eq:Delta}
\end{equation}
and
\begin{widetext}
\begin{equation}
\chi^{2}_{\rm{Lattice}}=\sum_{i,j=1}^{5}\left(f^{\rm{Lattice}}_{+,0}(q^{2})-P_{N}^{M}(q^{2})\right)_{i}({\rm{Cov}}_{ij}^{\rm{Lattice}})^{-1}\left(f^{\rm{Lattice}}_{+,0}(q^{2})-P_{N}^{M}(q^{2})\right)_{j}\,.\\[1ex]
\label{Eq:chi2LQCD}
\end{equation}
\end{widetext}

For the fit, we use the spectrum (and correlation) in 13 bins of $q^{2}$ $(N_{\rm{data}}=13)$ from the HFLAV group \cite{Amhis:2016xyh}, which results from the average of the four most precise measurements of the differential $B\to\pi\ell\nu_{\ell}$ decay rate from BaBar \cite{delAmoSanchez:2010af,Lees:2012vv} and Belle \cite{Ha:2010rf,Sibidanov:2013rkk}, the theoretical prediction of the partial decay rate Eq.\,(\ref{Eq:DecayDistribution}) and the $B^{0}$-meson lifetime $\tau_{B^{0}}$.
For the Lattice QCD information on the shape of the vector and scalar form factors, contained in $f^{\rm{Lattice}}_{+,0}(q^{2})$ in Eq.\,(\ref{Eq:chi2LQCD}), we use the results from the FLAG group \cite{Aoki:2019cca}, which are given in their Table\,(41).
However, these are presented as a formula, resulting from fits to a $z$-parametrization with 5 fit parameters, rather than as synthetic data for several values of $q^{2}$. 
For our analysis, we have generated synthetic data at three representative values of $q^{2}$ from their $z$-fits.
In particular, we have generated, respectively, 3 and 2 data points for the vector and scalar form factors $(N_{\rm{Lattice}}=5)$, which we gather in Table\,\ref{Table:GeneratedFLAGData} and use in our fits.\footnote{
Although synthetic data can be easily generated from the $z$-parametrization results, choosing the number of points and the $q^{2}$ leading to an optimal description of the form factors is not as straightforward.
In our case, we can generate 5 data points at most, as it would be inconsistent to generate more synthetic data than the independent coefficients of the $z$-fit; if more are generated, the resulting correlation matrix has zero eigenvalues, which implies a non-invertible covariance matrix.
We have checked that a $z$-fit with 5 parameters to the data given in Table\,\ref{Table:GeneratedFLAGData} yields the results of Table\,(41) from FLAG \cite{Aoki:2019cca}.
In our opinion, it would be more beneficial if the Lattice form factor calculations would be made available at some representative $q^{2}$ values along with the corresponding bin-to-bin correlation, apart from the parametrization coefficients of the $z$-fit, such that the results can be independently parametrized without assumptions on the functional form of the form factors.\label{footnote}} 

\begin{figure*}
\includegraphics[scale=0.425]{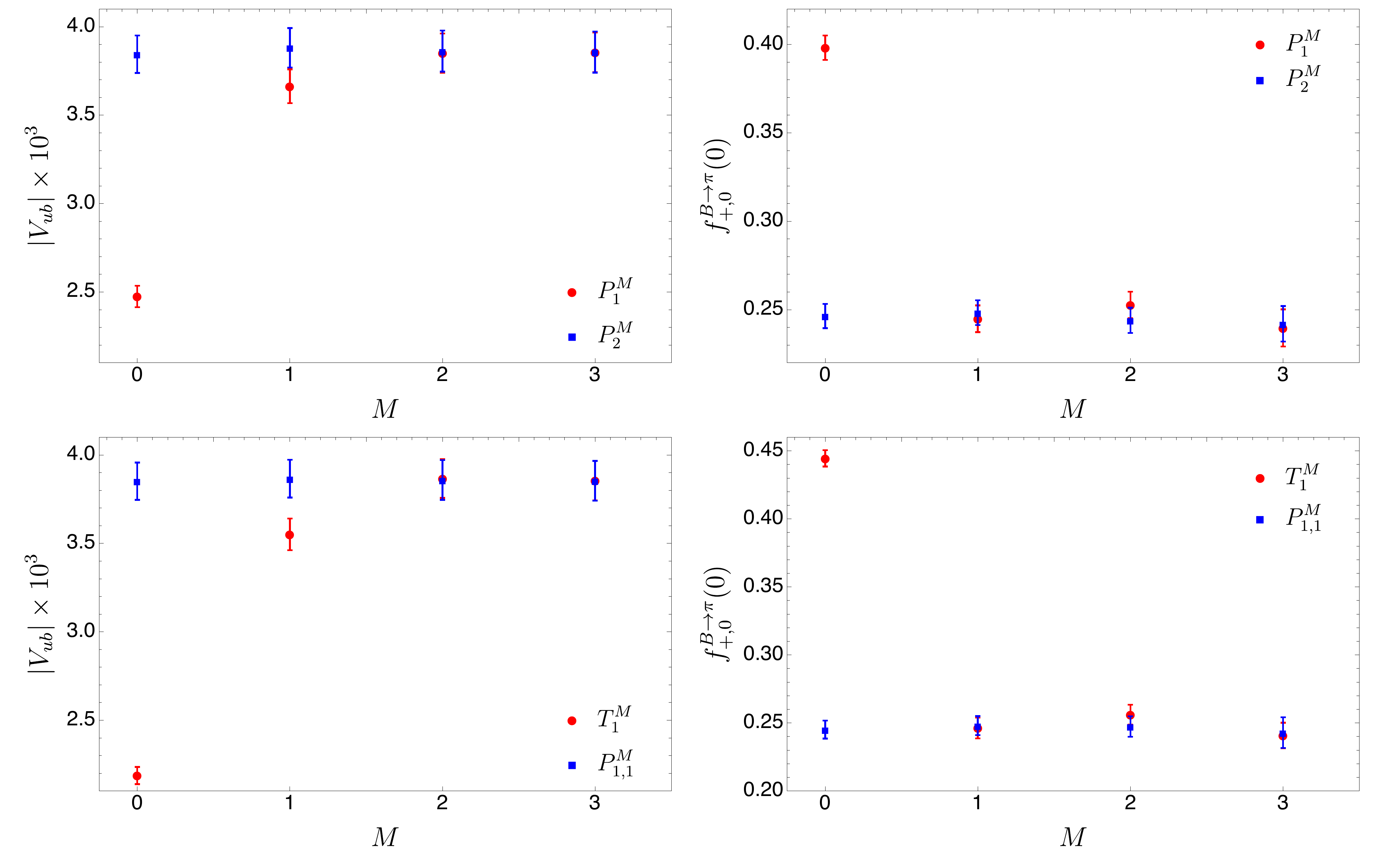}
\caption{Convergence pattern of the $P_{1}^{M}$ and $P_{2}^{M}$ (upper panels), and $T_{1}^{M}$ and $P_{1,1}^{M}$ (lower panels) sequences for $|V_{ub}|$ and $f_{+,0}^{B\to\pi}(0)$ resulting from our fits in Tables\,\ref{Table:FitSM} and \ref{Table:FitSMPadeType}.}
\label{Fig:PatternConvergenceBtoPi} 
\end{figure*}

\begin{figure*}
\includegraphics[width=8.75cm]{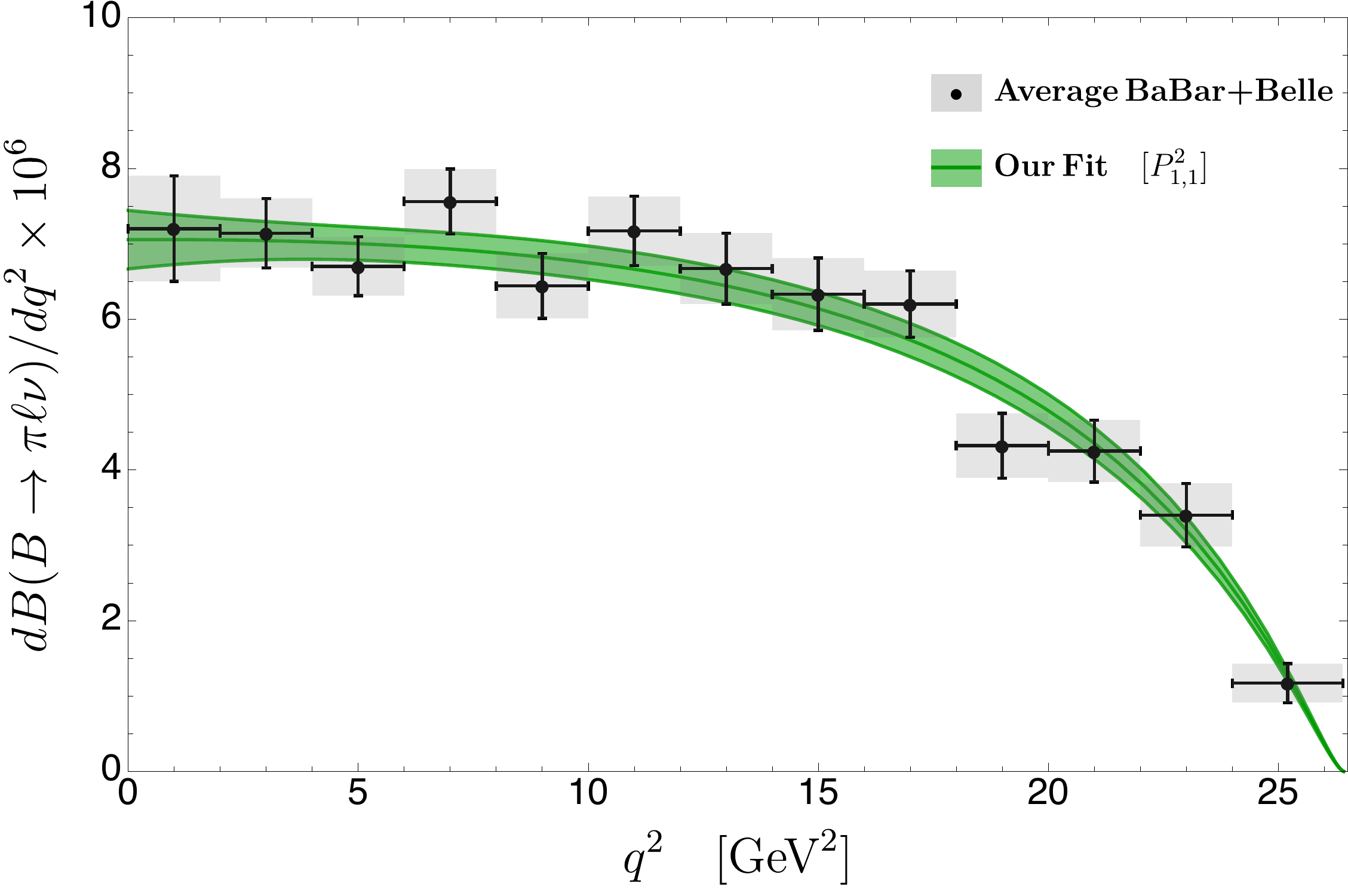}\quad\includegraphics[width=8.75cm]{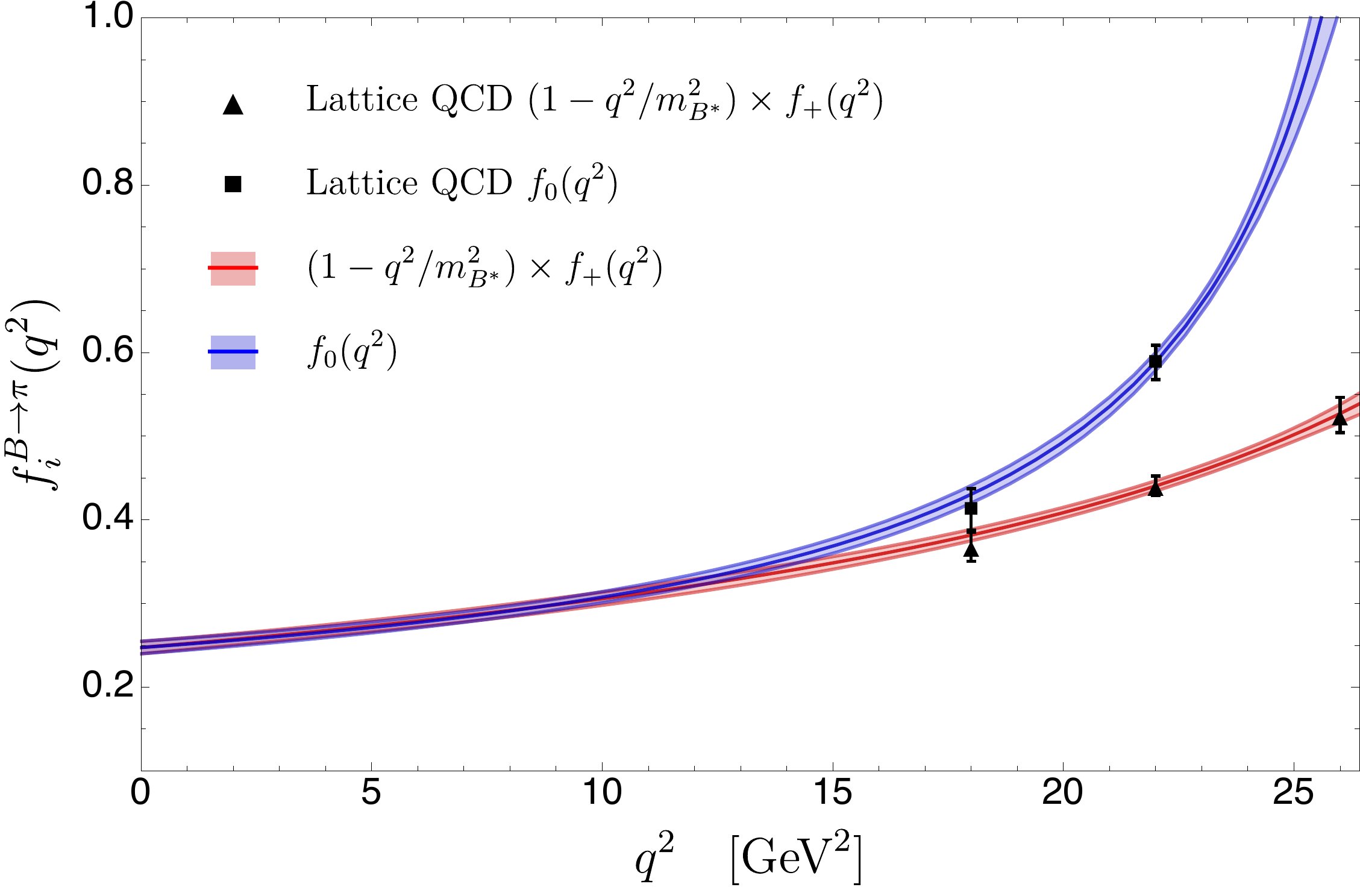}
\caption{{\it{Left}}: Averaged BaBar and Belle $B\to\pi\ell\nu$ differential branching ratio distribution (gray) \cite{Amhis:2016xyh} as compared to our $P_{1,1}^{2}$ result (green) obtained in combined fits as presented in Table\,\ref{Table:FitSMPadeType}. 
{\it{Right}}: Output for the $B\to\pi$ vector (red) and scalar (blue) form factors.}
\label{Fig:BPi} 
\end{figure*}

\begin{table*}
\begin{center}
\begin{tabular}{|cc|c|ccc|cc|}
\hline
&&&\multicolumn{5}{c|}{\multirow{1}{*}{Correlation matrix}}\\
\cline{4-8}
Form factor&&&\multicolumn{3}{c|}{\multirow{1}{*}{$f_{+}^{B\pi}$}} &\multicolumn{2}{c|}{\multirow{1}{*}{$f_{0}^{B\pi}$}} \\
&$q^{2}$ [GeV$^{2}$]&Central values&18&22&26&18&22\\
\hline
\multirow{3}{*}{$f_{+}^{B\pi}$}&18&1.007(48)&1&0.615&0.129&0.586&0.151\\
&22&1.967(52)&&1&0.382&0.170&0.245\\
&26&6.332(256)&&&1&0.306&0.221\\
\hline
\multirow{2}{*}{$f_{0}^{B\pi}$}&18&0.413(25)&&&&1&0.734\\
&22&0.588(21)&&&&&1\\
\hline
\end{tabular}
\caption{Central values, uncertainties and correlation matrix for the $B\to\pi$ vector and scalar form factors, $f^{B\to\pi}_{+,0}(q^{2})$, generated at three representative values of $q^{2}$ from the FLAG results \cite{Aoki:2019cca} and used in our fits in Eqs.\,(\ref{Eq:chi2}) and (\ref{Eq:chi2Combined}).}
\label{Table:GeneratedFLAGData}
\end{center}
\end{table*}
For the dominant vector form factor, we start fitting with Pad\'{e} sequences of the type $P_{1}^{M}(q^{2})$ and $P_{2}^{M}(q^{2})$, where the poles are left free to be fitted, and we reach, respectively, $M=3$ and $M=2$ as the best approximants with the current data.
The results of the fits for $|V_{ub}|$ and the fitted coefficients are presented in Table\,\ref{Table:FitSM} for the two Pad\'{e} sequences.\footnote{In the table, the element $P_{2}^{3}(q^{2})$ is only shown for illustration.}
In the table, the poles denoted by the symbol $\dagger$ are Froissart doublet poles.\footnote{The element $P_{2}^{2}(q^{2})$ (also the $P_{2}^{3}(q^{2})$) has complex-conjugate poles with an small imaginary which are pair up by a close-by zero in the numerator, thus becoming effectively a defect, also called Froissart doublet. 
These poles lie within the radius of convergence, indicating certain degree of unitarity violation in the data \cite{Gonzalez-Solis:2018ooo}, since their presence is forbidden when dealing with Stieltjes functions.}
We also show the coefficients of the $P_{1}^{1}(q^{2})$ approximants used for the description of the scalar form factor, which provides an optimal description of the data.\footnote{We have also tried a $P_{1}^{2}(q^{2})$ approximant for the scalar form factor and found no impact on $|V_{ub}|$.}
The latter contains only 2 free parameters, $a_{1}^{0}$ and the effective $m_{B^{*}(0^{+})}$ pole, as in our fits the constraint at $q^{2}=0$, {\it{i.e.}} $f_{+}^{B\to\pi}(0)=f_{0}^{B\to\pi}(0)$ (cf.\,Eq.\,(\ref{Eq:FormFactorConstraint})), has been implemented explicitly through $a_{0}^{+}=a_{0}^{0}$.
Had we fit with sequences of the type $T_{1}^{M}(q^{2})$ and $P_{1,1}^{M}(q^{2})$, where the $B^{*}(1^{-})$ pole is fixed to the PDG mass, $m_{B^{*}(1^{-})}=5.325$ GeV \cite{Zyla:2020zbs}, we would have reached, respectively, $M=3$ and $M=2$ as the best approximants and obtained the results collected in Table\,\ref{Table:FitSMPadeType}.
In Fig.\,\ref{Fig:PatternConvergenceBtoPi} we provide a graphical account of the convergence pattern for $|V_{ub}|$ and $f_{+,0}^{B\to\pi}(0)$ resulting from the four types of sequences we have considered.
The stability observed for these quantities is quite reassuring.
The values obtained for the individual $\chi^{2}$ functions, $\chi^{2}_{\rm{data}}$ and $\chi^{2}_{\rm{Lattice}}$, imply a good quality of the fits.
Furthermore, we note that the approximants with two poles yield excellent values for the quantity $(\chi^{2}_{\rm{data}}+\chi^{2}_{\rm{Lattice}})/{\rm{dof}}$.
In terms of the latter, our best fit\footnote{Our best fit is defined as the last approximant of a given sequence with all parameters different from zero at one-sigma distance and with $\chi^2/dof$ closer to $1$} is obtained with a $P_{1,1}^{2}$ approximant, which yields:
\begin{equation} 
|V_{ub}|=3.86(11)\times10^{-3}\,,
\label{Eq:BestFitBtoPi}
\end{equation}
although the values of $|V_{ub}|$ obtained with the other approximants are almost identical as it can be seen on the tables.
For our best fit, $P_{1,1}^{2}$, the quoted uncertainty on $|V_{ub}|$ is $2.9\%$ (cf.\,Eq.\,(\ref{Eq:BestFitBtoPi})) and we gather the resulting fit parameters along with the correlation matrix in Table \ref{Table:CorrelationBestFitBtoPi} of Appendix \ref{Appendix1}.
Our $|V_{ub}|$ value in Eq.\,(\ref{Eq:BestFitBtoPi}) is larger, and slightly more precise than, the FNAL/MILC result, $|V_{ub}|=3.72(16)\times10^{-3}$ \cite{Lattice:2015tia}, and the FLAG reported value, $|V_{ub}|=3.73(14)\times10^{-3}$ \cite{Aoki:2019cca}.
The reason for that is due to the adopted $\chi^{2}$ fit function in Eq.\,(\ref{Eq:chi2}), which we consider as more democratic.
In addition, this procedure has an impact on the comparison with respect to $|V_{ub}|$ determinations from inclusive decays $B\to X_{u}\ell\nu_{\ell}$, $|V_{ub}|=4.25(12)^{+15}_{-14}(23)\times10^{-3}$ \cite{Zyla:2020zbs}, with which our values differ by only $1.35\sigma$.
In Fig.\,\ref{Fig:BPi}, we show the differential branching ratio distribution (left plot) and the outputs for the vector and scalar form factors (right plot) resulting from our preferred fit $P_{1,1}^{2}$.
\begin{table*}
\begin{center}
\begin{tabular}{|l|cccc|cccc|}
\hline
&\multicolumn{4}{c|}{Element of the $P_{1}^{M}$ sequence}&\multicolumn{4}{c|}{Element of the $P_{2}^{M}$ sequence}\\
\cline{2-9}
Parameter&$P_{1}^{0}$&$P_{1}^{1}$&$P_{1}^{2}$&$P_{1}^{3}$&$P_{2}^{0}$&$P_{2}^{1}$&$P_{2}^{2}$&$P_{2}^{3}$\\
\hline
$|V_{ub}|\times10^{3}$ &$2.47(6)$&$3.66(10)$&$3.85(11)$&$3.86(11)$&$3.85(11)$&$3.88(11)$&$3.86(12)$&$3.86(12)$\\ 
$a_{0}^{+}$ &$0.398(7)$&$0.245(8)$&$0.253(8)$&$0.240(11)$&$0.246(7)$&$0.248(7)$&$0.244(7)$&$0.242(10)$\\ 
$a_{1}^{+}\times10^{3}$ &---&$7.9(4)$&$2.8(1.4)$&$8.1(3.3)$&---&$-1.9(1.4)$&$-3.5(9)$&$-2.5(4.5)$\\ 
$a_{2}^{+}\times10^{4}$ &---&---&$2.4(6)$&$-3.3(3.3)$&---&---&$-1.7(8)$&$-2.5(2.4)$\\ 
$a_{3}^{+}\times10^{5}$ &---&---&---&$1.7(1.0)$&---&---&---&$0.2(9)$\\ 
$m_{B^{*}(1^{-})}$ pole(s) [GeV] &$5.26$&$5.29$&$5.31$&$5.33$&$5.32\&7.11$&$5.34\&6.40$&$\dagger$&$\dagger$\\ 
\hline
$a_{1}^{0}\times10^{2}$ &$-1.3(1)$&$-0.2(1)$&$-0.5(1)$&$-0.4(1)$&$-0.4(1)$&$-0.5(1)$&$-0.5(1)$&$-0.5(1)$\\ 
$m_{B^{*}(0^{+})}$ pole [GeV] &$5.17$&$5.72$&$5.45$&$5.43$&$5.47$&$5.39$&$5.38$&$5.38$\\ 
\hline
$\chi^{2}_{\rm{data}}\,[N_{\rm{data}}=13]$&$157.07$&$12.64$&$11.51$&$11.92$&$10.76$&$11.87$&$10.80$&$10.90$\\
$\chi^{2}_{\rm{Lattice}}\,[N_{\rm{Lattice}}=5]$&$18.19$&$5.15$&$1.72$&$0.67$&$1.53$&$0.75$&$0.42$&$0.34$\\ 
$(\chi^{2}_{\rm{data}}+\chi^{2}_{\rm{Lattice}})/{\rm{dof}}$&$13.48$&$1.48$&$1.20$&$1.26$&$0.95$&$1.05$&$1.02$&$1.12$\\ 
\hline
\end{tabular}
\caption{Best fit values and uncertainties for the output quantities of our $\chi^{2}_{B\pi}$ fits Eq.\,(\ref{Eq:chi2}) for Pad\'{e} sequences of the type $P_{1}^{M}$ and $P_{2}^{M}$.
}
\label{Table:FitSM}
\end{center}
\end{table*}

\begin{table*}
\begin{center}
\begin{tabular}{|l|cccc|cccc|}
\hline
&\multicolumn{4}{c|}{Element of the $T_{1}^{M}$ sequence}&\multicolumn{4}{c|}{Element of the $P_{1,1}^{M}$ sequence}\\
\cline{2-9}
Parameter&$T_{1}^{0}$&$T_{1}^{1}$&$T_{1}^{2}$&$T_{1}^{3}$&$P_{1,1}^{0}$&$P_{1,1}^{1}$&$P_{1,1}^{2}$&$P_{1,1}^{3}$\\
\hline
$|V_{ub}|\times10^{3}$ &$2.19(5)$&$3.55(9)$&$3.87(11)$&$3.85(11)$&$3.85(11)$&$3.87(11)$&$3.86(11)$&$3.85(11)$\\ 
$a_{0}^{+}$ &$0.445(6)$&$0.246(8)$&$0.256(7)$&$0.241(9)$&$0.245(7)$&$0.248(7)$&$0.247(8)$&$0.243(11)$\\ 
$a_{1}^{+}\times10^{3}$ &---&$9.1(3)$&$1.5(1.2)$&$7.7(2.7)$&---&$-1.3(9)$&$-1.3(8)$&$3.5(11.4)$\\ 
$a_{2}^{+}\times10^{4}$ &---&---&$3.2(5)$&$-2.7(2.3)$&---&---&$-0.3(1.0)$&$-1.9(3.3)$\\ 
$a_{3}^{+}\times10^{5}$ &---&---&---&$1.5(6)$&---&---&---&$0.9(2.0)$\\ 
$m_{B^{*}(1^{-})}$ pole(s) [GeV] &$=5.325$&$=5.325$&$=5.325$&$=5.325$&$=5.325\&7.03$&$=5.325\&6.64$&$=5.325\&6.46$&$=5.325\&8.97$\\ 
\hline
$a_{1}^{0}\times10^{2}$ &$-1.9(1)$&$-0.4(1)$&$-0.5(1)$&$-0.4(1)$&$-0.4(1)$&$-0.5(1)$&$-0.4(1)$&$-0.4(1)$\\
$m_{B^{*}(0^{+})}$ pole [GeV] &$4.78$&$5.57$&$5.36$&$5.44$&$5.45$&$5.43$&$5.44$&$5.44$\\ 
\hline
$\chi^{2}_{\rm{data}}\,[N_{\rm{data}}=13]$&$182.19$&$17.21$&$13.64$&$11.65$&$11.27$&$11.26$&$10.95$&$11.17$\\
$\chi^{2}_{\rm{Lattice}}\,[N_{\rm{Lattice}}=5]$&$41.05$&$11.53$&$1.93$&$0.78$&$1.57$&$1.04$&$1.15$&$0.92$\\ 
$(\chi^{2}_{\rm{data}}+\chi^{2}_{\rm{Lattice}})/{\rm{dof}}$&$15.95$&$2.21$&$1.30$&$1.13$&$0.92$&$0.95$&$1.01$&$1.10$\\ 
\hline
\end{tabular}
\caption{Best fit values and uncertainties for the output quantities of our $\chi^{2}_{B\pi}$ fits Eq.\,(\ref{Eq:chi2}) for Pad\'{e} sequences of the type $T_{1}^{M}$ and $P_{1,1}^{M}$.}
\label{Table:FitSMPadeType}
\end{center}
\end{table*}

Had we performed an analysis including only the vector form factor Lattice data into the fit,\footnote{For this fit, we have taken the limit $m_{\ell}\to0$ in Eq.\,(\ref{Eq:DecayDistribution}) and used the synthetic data from Table \ref{Table:CoefficientsVectorAlone} of Appendix \ref{Appendix1}, which have been generated from the FLAG standalone $z$-fit to the vector form factor given in Eq.\,(224) in \cite{Aoki:2019cca}.} 
we would have reached $M=2$ and obtained the results shown in Table \ref{Table:FitSMvectorSequence}.\footnote{As a matter of example, in this table we only report $P_{1}^{M}$ approximants. 
Similar results and conclusions are obtained using the other approximants considered in Tables \ref{Table:FitSM} and \ref{Table:FitSMPadeType}.}
Note that $|V_{ub}|$ in this fit, $|V_{ub}|=3.65(11)\times10^{-3}$, shifts by about $\sim1.3\sigma$ downwards with respect to the value given in Eq.\,(\ref{Eq:BestFitBtoPi}), $|V_{ub}|=3.86(11)\times10^{-3}$, obtained with the scalar form factor Lattice data taken into account.
The origin of this shift stems from the fact that the FLAG value for $f_{+}^{B\to\pi}(0)$ resulting from a standalone $z$-fit to the vector form factor, $f_{+}^{B\to\pi}(0)=0.288(87)$ \cite{Aoki:2019cca}, which is the most relevant input for the extraction of $|V_{ub}|$, shifts by about $1.2\sigma$ upwards with respect to their $z$-fits including the scalar form factor, $f_{+}^{B\to\pi}(0)=0.139(90)$ \cite{Aoki:2019cca}, which is obtained with the restriction $f_{+}^{B\to\pi}(0)=f_{0}^{B\to\pi}(0)$.
In this case, our $|V_{ub}|$ value is found to be in line with the HFLAV result, $|V_{ub}|=3.70(10)(12)\times10^{-3}$ \cite{Amhis:2016xyh}, obtained from $z$-fits with the vector form factor only; our central value is slightly smaller due to the form adopted in Eq.\,(\ref{Eq:chi2}).

\begin{table*}
\begin{center}
\begin{tabular}{|l|cccc|}
\hline
&\multicolumn{4}{c|}{Element of the Pad\'{e} sequence}\\
\cline{2-5}
Parameter&$P_{1}^{0}$&$P_{1}^{1}$&$P_{1}^{2}$&$P_{1}^{3}$\\
\hline
$|V_{ub}|\times10^{3}$ &$2.40(6)$&$3.56(9)$&$3.65(11)$&$3.66(11)$\\ 
$a_{0}^{+}$ &$0.409(6)$&$0.251(8)$&$0.256(8)$&$0.260(11)$\\ 
$a_{1}^{+}\times10^{3}$ &---&$8.3(4)$&$5.8(1.4)$&$3.5(3.5)$\\ 
$a_{2}^{+}\times10^{4}$ &---&---&$1.2(7)$&$3.5(3.3)$\\ 
$a_{3}^{+}\times10^{6}$ &---&---&---&$-6.6(9.4)$\\ 
$m_{B^{*}(1^{-})}$ pole [GeV] &$5.28$&$5.31$&$5.33$&$5.32$\\ 
\hline
$\chi^{2}_{\rm{data}}\,[N_{\rm{data}}=13]$&$163.01$&$14.82$&$11.80$&$11.84$\\
$\chi^{2}_{\rm{Lattice}}\,[N_{\rm{Lattice}}=3]$&$5.80$&$0.004$&$0.16$&$0.05$\\ 
$(\chi^{2}_{\rm{data}}+\chi^{2}_{\rm{Lattice}})/{\rm{dof}}$&$11.25$&$1.06$&$0.92$&$0.99$\\ 
\hline
\end{tabular}
\caption{Best fit values, uncertainties and correlation matrix for the output quantities of our $\chi^{2}_{B\pi}$ fits Eq.\,(\ref{Eq:chi2}) obtained from the averaged $B\to\pi\ell\nu_{\ell}$ BaBar and Belle experimental data \cite{Amhis:2016xyh} in combination with the Lattice-QCD vector form factor simulations \cite{Aoki:2016frl}.}
\label{Table:FitSMvectorSequence}
\end{center}
\end{table*}

\subsection{Fits to the decay $B_{s}\to K\ell\nu_{\ell}$}\label{section3sub2}

For the determination of $|V_{ub}|$ from the decay $B_{s}\to K\ell\nu_{\ell}$, we follow a strategy similar to that of the previous section for $B\to\pi\ell\nu$, using recent experimental information on the decay spectrum together with the form factors shape information from theory given by the Lattice-QCD Collaborations.

The RBC/UKQCD Lattice Collaboration provides its results for both the vector and scalar form factors as synthetic, correlated data at three representative $q^{2}$ values in Tables VI and IX of Ref.\,\cite{Flynn:2015mha},
while the FNAL/MILC Lattice Collaboration presents their as a formula resulting from fits to a $z$-expansion parametrization with 8 fit coefficients, which are given in Table X of Ref.\,\cite{Bazavov:2019aom}.
For our study, we have generated synthetic data of the latter at four representative values of $q^{2}$ from their $z$-fits.
In particular, we have generated 4 and 3 data points for the vector and scalar form factors, respectively, which we collect in Table \ref{Table:GeneratedFNALMILCData}.\footnote{At most, we can generate 7 data points, as it would be inconsistent to generate more data than the independent coefficients of the $z$-fit; if more are generated, the resulting covariance matrix is not invertible.}
We will next use these results, which can combined with the binned branching ratio LHCb measurements, $BR(B_{s}\to K^{-}\mu^{+}\nu_{\mu})=0.36(2)(3)\times10^{-4}$ for $q^{2}<7$ GeV$^{2}$ and $BR(B_{s}\to K^{-}\mu^{+}\nu_{\mu})=0.70(5)(6)\times10^{-4}$ for $q^{2}>7$ GeV$^{2}$ \cite{Aaij:2020nvo}, to determine $|V_{ub}|$.

\begin{table*}
\begin{center}
\begin{tabular}{|cc|c|cccc|ccc|}
\hline
&&&\multicolumn{7}{c|}{\multirow{1}{*}{Correlation matrix}}\\
\cline{4-10}
Form factor&&&\multicolumn{4}{c|}{\multirow{1}{*}{$f_{+}^{B_{s}K}$}} &\multicolumn{3}{c|}{\multirow{1}{*}{$f_{0}^{B_{s}K}$}} \\
&$q^{2}$ [GeV$^{2}$]&Central values&17&19&21&23&17&19&21\\
\hline
\multirow{4}{*}{$f_{+}^{B_{s}K}$}&17&0.9268(428)&1&0.9572&0.7571&0.3615&0.6943&0.6749&0.5862\\
&19&1.2460(441)&&1&0.9096&0.5890&0.5778&0.6214&0.6071\\
&21&1.7530(516)&&&1&0.8653&0.3985&0.5057&0.5726\\
&23&2.6593(820)&&&&1&0.1885&0.3161&0.4235\\
\hline
\multirow{3}{*}{$f_{0}^{B_{s}K}$}&17&0.4219(196)&&&&&1&0.9499&0.7716\\
&19&0.4991(153)&&&&&&1&0.9267\\
&21&0.5974(136)&&&&&&&1\\
\hline
\end{tabular}
\caption{Central values, uncertainties and correlation matrix for the $B_{s}\to K$ vector and scalar form factors, $f^{B_{s}\to K}_{+,0}(q^{2})$, generated at four representative values of $q^{2}$ from the FNAL/MILC results \cite{Bazavov:2019aom} and used in our fits in Eqs.\,(\ref{Eq:chi2Bs}) and (\ref{Eq:chi2Combined}).}
\label{Table:GeneratedFNALMILCData}
\end{center}
\end{table*}

The form of the $\chi^{2}$ function to be minimized, analogous to that of Eq.\,(\ref{Eq:chi2}) for $B\to\pi$, is given by:
\begin{eqnarray}
\chi^{2}_{B_{s}K}=N\left(\frac{\chi^{2}_{\rm{LHCb}}}{N_{\rm{LHCb}}}+\frac{\chi^{2}_{\rm{RBC/UKQCD}}}{N_{\rm{RBC/UKQCD}}}+\frac{\chi^{2}_{\rm{FNAL/MILC}}}{N_{\rm{FNAL/MILC}}}\right)\,,
\label{Eq:chi2Bs}
\end{eqnarray}
where $N_{\rm{LHCb}}=2$ is the number of experimental points, while $N_{\rm{RBC/UKQCD}}=6$ and $N_{\rm{FNAL/MILC}}=7$ are the number of the RBC/UKQCD
and FNAL/MILC Lattice points, respectively, and $N=N_{\rm{LHCb}}+N_{\rm{RBC/UKQCD}}+N_{\rm{FNAL/MILC}}$.
The first term in Eq.\,(\ref{Eq:chi2Bs}),
\begin{equation}
\chi^{2}_{\rm{LHCb}}=\sum_{i=1}^{2}({BR}^{\rm{exp}}_{i}-{BR}^{\rm{th}}_{i})^{2}/\sigma_{{BR}^{\rm{exp}}_{i}}^{2}\,,
\end{equation}
contains the information of the LHCb experimental measurements of the branching ratio in the (uncorrelated) low and high $q^{2}$ regions, ${BR}^{\rm{exp}}_{i}$ is the measured branching ratio and $\sigma_{BR_{i}}^{\rm{exp}}$ the corresponding uncertainty in the $i$-th bin, while the second and third terms include the theoretical information on the form factors from Lattice through a $\chi^{2}$ function of the form:
\begin{widetext}
\begin{equation}
\chi^{2}_{\rm{Lattice}}=\sum_{i,j=1}^{N_{\rm{Lattice}}}\left(f_{+,0}^{{\rm{Lattice}}}(q^{2})-f_{+,0}(q^{2})\right)_{i}\left({\rm{Cov}}_{ij}^{\rm{Lattice}}\right)^{-1}\left(f_{+,0}^{{\rm{Lattice}}}(q^{2})-f_{+,0}(q^{2})\right)_{j}\,.
\label{Eq:Latticechi2}
\end{equation}
\end{widetext}

Table \ref{Table:FitDataAllLatticeBs} summarizes the best fit values for $|V_{ub}|$ and the form factor parameters for the various Pad\'{e} sequences.
These fits have been performed using a $P_{1}^{0}$ approximant for the scalar form factor and taking the $f_{+}^{B_{s}\to K}(0)=f_{0}^{B_{s}\to K}(0)$ restriction into account (cf.\,Eq.\,(\ref{Eq:FormFactorConstraint})), thus having the $m_{B^{*}(0^{+})}$ pole as the only free parameter in the scalar sector.\footnote{We have also tried $P_{1}^{1}$ and $P_{1}^{2}$ approximants for the scalar form factors and found that the fit parameters remain stable.} 
The values of the $\chi^{2}$ functions reported in the tables imply a very good quality of the fits.
For the single pole Pad\'{e} sequences $P_{1}^{M}$ and $T_{1}^{M}$, we find the fits stabilize for $M=3$ and the obtained $|V_{ub}|$ value, $|V_{ub}|=3.58(8)\times10^{-3}$, has an uncertainty of $2.2\%$.
For the sequences with two poles, we reach $P_{2}^{2}$ and $P_{1,1}^{3}$ and obtain $|V_{ub}|=3.60(9)\times10^{-3}$ and $|V_{ub}|=3.58(9)\times10^{-3}$, respectively, which is a $2.5\%$ error.
As seen, the values for $|V_{ub}|$ obtained with the various approximants are almost identical.
In terms of the quantity $(\chi^{2}_{\rm{LHCb}}+\chi^{2}_{\rm{RBC/UKQCD}}+\chi^{2}_{\rm{FNAL/MILC}})/$dof, the approximants $P_{1}^{3}$ and $P_{1,1}^{3}$ yield the best fits.\footnote{Note that the second pole of the approximant $P_{1,1}^{3}$ is placed far away from the origin and it thus behaves as a $P_{1}^{3}$.} 
These values for $|V_{ub}|$ represent a shift of about $(1.8-2)\sigma$ downwards with respect to the value $|V_{ub}|=3.86(11)\times10^{-3}$ determined from the decay $B\to\pi\ell\nu_{\ell}$ (cf.\,Eq.\,(\ref{Eq:BestFitBtoPi})).
Despite the differing results, we note that an important aspect to improve the compatibility results for $|V_{ub}|$ is the binned measurement of the $B_{s}\to K\ell\nu_{\ell}$ differential branching ratio distribution, and most importantly its low-energy region, which fixes the $q^{2}$-dependence of the form factors at low-energies.
In this sense, the experimental information is presently limited to the two LHCb experimental points, which are rather thick for an accurate extraction of the functional behavior of the form factors, specially at low-energies. 
Therefore, new and more precise measurements of the decay rate with a thinner resolution of the $q^{2}$ bins will definitely allow obtain more conclusive results from the $B_{s}\to K\ell\nu_{\ell}$ decay.

\begin{figure*}
\includegraphics[width=8.75cm]{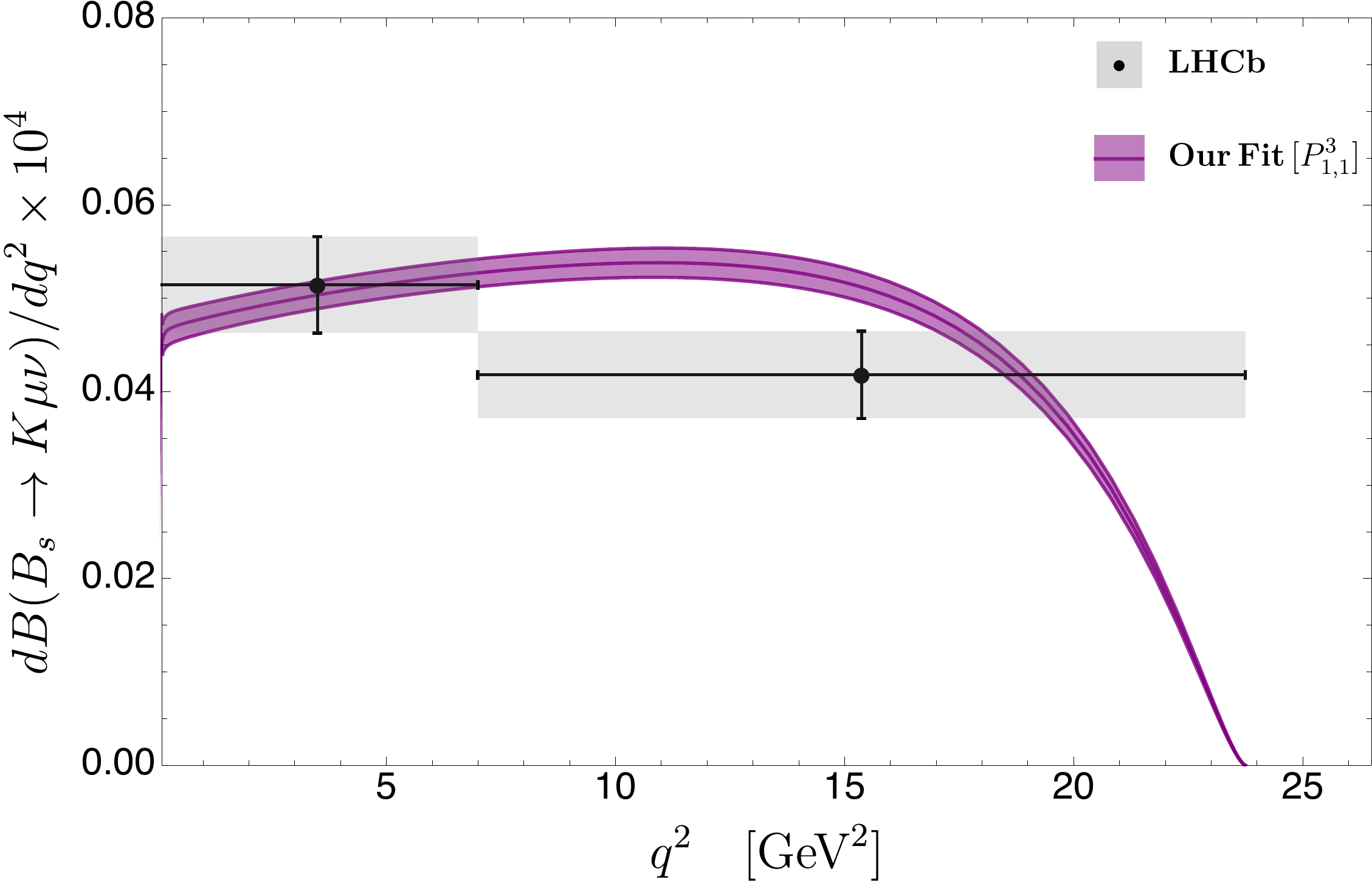}\quad\includegraphics[width=8.75cm]{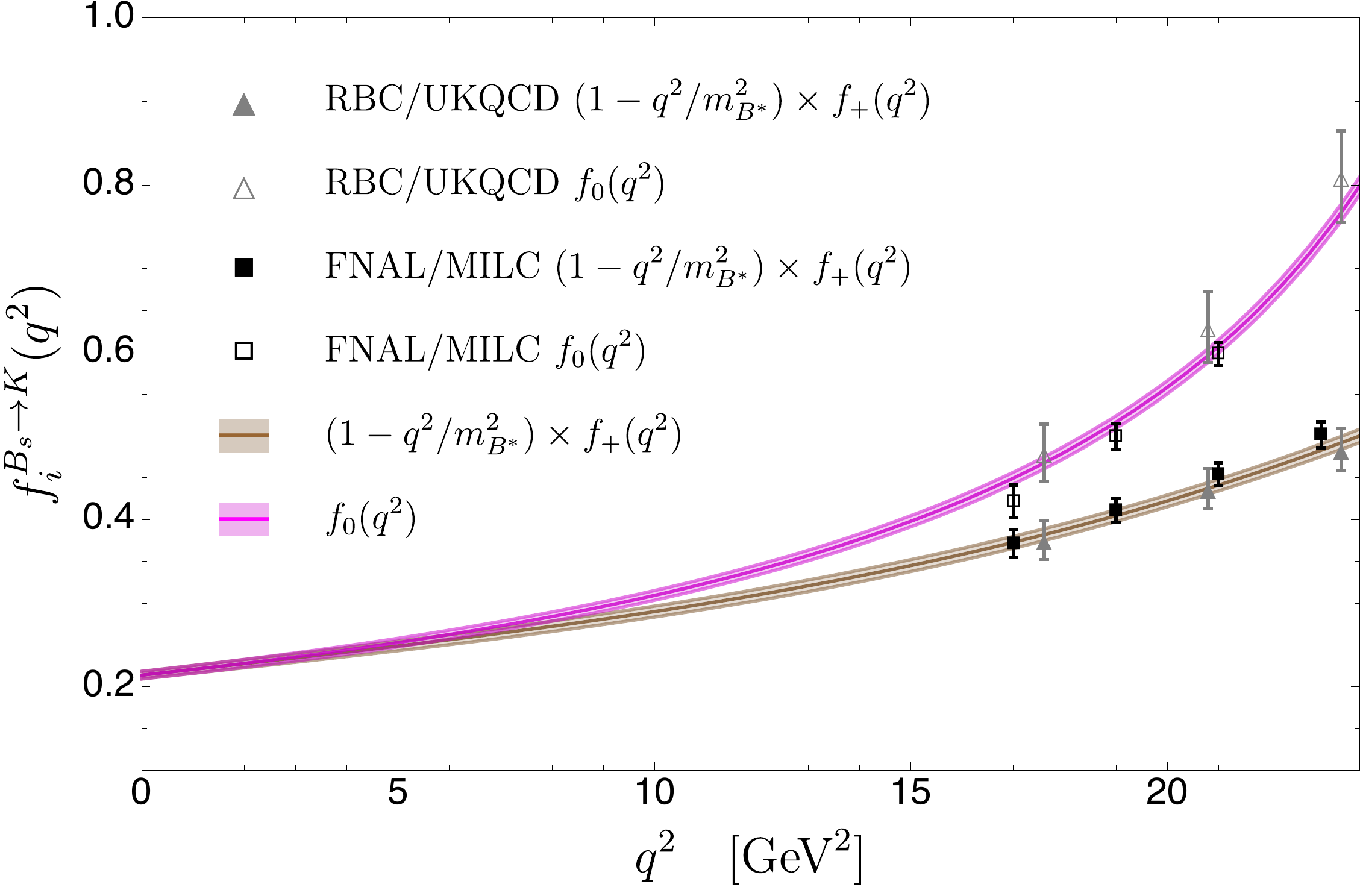}
\caption{{\it{Left}}: LHCb $B_{s}\to K^{-}\mu^{+}\nu_{\mu}$ differential branching ratio distribution (gray) \cite{Aaij:2020nvo} as compared to our best fit result (purple) obtained in combined fits as presented in Table\,\ref{Table:FitDataAllLatticeBs}; the two LHCb data points are placed in the middle of each bin and have been divided by the bin width.
{\it{Right}}: Output for the $B_{s}\to K$ vector (brown) and scalar (magenta) form factors compared to the Lattice-QCD data of Ref.\,\cite{Flynn:2015mha} and Table \ref{Table:GeneratedFNALMILCData}.}
\label{Fig:Bs} 
\end{figure*}

A graphical account of our fit with the $P_{1,1}^{3}$ approximant is presented in Fig.\,\ref{Fig:Bs} for the differential branching ratio distribution (left plot) and the output for the vector and scalar form factors (right plot), while the resulting parameters and correlation matrix of this fit is given Table \ref{Table:CorrelationBestFitBstoK} of Appendix \ref{Appendix1}.
\begin{table*}
\begin{center}
\begin{tabular}{|l|cccc|cccc|}
\hline
&\multicolumn{4}{c|}{Pad\'{e} element}\\
\cline{2-5}
Parameter&$P_{1}^{3}$&$P_{2}^{2}$&$T_{1}^{3}$&$P_{1,1}^{3}$\\
\hline
$|V_{ub}|\times10^{3}$ &$3.58(8)$&$3.60(9)$&$3.58(8)$&$3.58(9)$\\ 
$a_{0}^{+}$ &$0.214(5)$&$0.214(5)$&$0.214(5)$&$0.214(5)$\\ 
$a_{1}^{+}\times10^{3}$ &$7.02(40)$&$1.12(65)$&$7.02(40)$&$6.70(5.40)$\\ 
$a_{2}^{+}\times10^{4}$ &$-0.55(23)$&$0.16(20)$&$-0.50(14)$&$-0.48(46)$\\ 
$a_{3}^{+}\times10^{5}$ &$1.12(14)$&---&$1.10(13)$&$1.04(96)$\\ 
$m_{B^{*}(1^{-})}$ pole(s) [GeV] &$5.32$&$5.33\&6.83$&$=5.325$&$=5.325\&29.5$\\ 
$m_{B^{*}(0^{+})}$ pole [GeV] &$5.70$&$5.69$&$5.70$&$5.70$\\ 
\hline
$\chi^{2}_{\rm{LHCb}}\,[N_{\rm{LHCb}}=2]$&$0.14$&$0.20$&$0.14$&$0.15$\\
$\chi^{2}_{\rm{RBC/UKQCD}}\,[N_{\rm{RBC/UKQCD}}=6]$&$3.25$&$3.17$&$3.21$&$3.21$\\ 
$\chi^{2}_{\rm{FNAL/MILC}}\,[N_{\rm{FNAL/MILC}}=7]$&$4.89$&$5.00$&$4.95$&$4.94$\\ 
$(\chi^{2}_{\rm{LHCb}}+\chi^{2}_{\rm{RBC/UKQCD}}+\chi^{2}_{\rm{FNAL/MILC}})/{\rm{dof}}$&$1.03$&$1.05$&$0.92$&$1.03$\\ 
\hline
\end{tabular}
\caption{Best fit values and uncertainties for the output quantities of our $\chi^{2}_{B_{s}K}$ fits Eq.\,(\ref{Eq:chi2Bs}) for the various Pad\'{e} sequences.
}
\label{Table:FitDataAllLatticeBs}
\end{center}
\end{table*}

\subsection{Combined fits to the decays $B\to\pi\ell\nu_{\ell}$ and $B_{s}\to K\ell\nu_{\ell}$}\label{section3sub3}

In the previous Secs. \ref{section3sub1} and \ref{section3sub2} we have extracted $|V_{ub}|$ and the corresponding form factor parameters from individual fits to the decays $B\to\pi\ell\nu_{\ell}$ and $B_{s}\to K^{-}\mu^{+}\nu_{\mu}$ experimental data combined with the Lattice-QCD information on the corresponding vector and scalar form factors.
In this section, we explore the potential of performing simultaneous fits to all experimental and theoretical information on both exclusive decays to determine $|V_{ub}|$.
For that, we proceed in a similar fashion as in the previous cases, Eqs.\,(\ref{Eq:chi2}) and (\ref{Eq:chi2Bs}), and minimize the following $\chi^{2}$-function: 
\begin{widetext}
\begin{equation}
\chi^{2}=N\left(\frac{\chi^{2}_{\rm{BaBar+Belle}}}{N_{\rm{BaBar+Belle}}}+\frac{\chi^{2}_{\rm{FLAG}}}{N_{\rm{FLAG}}}+\frac{\chi^{2}_{\rm{LHCb}}}{N_{\rm{LHCb}}}+\frac{\chi^{2}_{\rm{RBC/UKQCD}}}{N_{\rm{RBC/UKQCD}}}+\frac{\chi^{2}_{\rm{FNAL/MILC}}}{N_{\rm{FNAL/MILC}}}\right)\,,
\label{Eq:chi2Combined}
\end{equation}
\end{widetext}
where the first two terms contain the information on the decay $B\to\pi\ell\nu_{\ell}$ channel, while the three other include that of the $B_{s}\to K^{-}\mu^{+}\nu_{\mu}$ channel, with $N_{\rm{BaBar+Belle}}=13, N_{\rm{FLAG}}=5, N_{\rm{LHCb}}=2, N_{\rm{RBC/UKQCD}}=6, N_{\rm{FNAL/MILC}}=7$ and $N=N_{\rm{BaBar+Belle}}+N_{\rm{FLAG}}+N_{\rm{LHCb}}+N_{\rm{RBC/UKQCD}}+N_{\rm{FNAL/MILC}}$.
This definition equally weight each data set and prevents sets with a smaller data points, such as the $B_{s}\to K\ell\nu_{\ell}$ spectra, from being dominated by sets with a larger data points, such as the $B\to\pi\ell\nu_{\ell}$ spectra.

As in the preceding sections, we have tried various Pad\'{e} sequences. 
Here, however, we only show our results for $|V_{ub}|$ and the form factor parameters resulting from the partial Pad\'{e} sequence $P_{1,1}^{M}$, which yielded the best fit results in our previous individual analyses.
We reach $M=2$ and $M=3$ for the $B\to\pi$ and $B_{s}\to K$ vector form factors, respectively.
The resulting fit parameters and the correlation matrix are presented in Table \ref{Table:CombinedFit}\footnote{In the table, we use $c_{i}$ to denote the Pad\'{e} approximant fit parameters of the $B_{s}\to K$ form factors.}, which have been obtained taking into account the restrictions $f_{+}^{B\to\pi}(0)=f_{0}^{B\to\pi}(0)$ and $f_{+}^{B_{s}\to K}(0)=f_{0}^{B_{s}\to K}(0)$ simultaneously.
The value for the quantity $(\chi^{2}_{\rm{BaBar+Belle}}+\chi^{2}_{\rm{FLAG}}+\chi^{2}_{\rm{LHCb}}+\chi^{2}_{\rm{RBC/UKQCD}}+\chi^{2}_{\rm{FNAL/MILC}})/$dof$=1.08$ indicates a good quality of the fit.
The resulting value for $|V_{ub}|$ from the combined analysis is found to be:
\begin{equation}
|V_{ub}|=3.68(5)\times10^{-3}\,,
\label{Eq:VubCombined}
\end{equation}
which is only a $1.4\%$ error.

\begin{figure*}
\includegraphics[width=8.75cm]{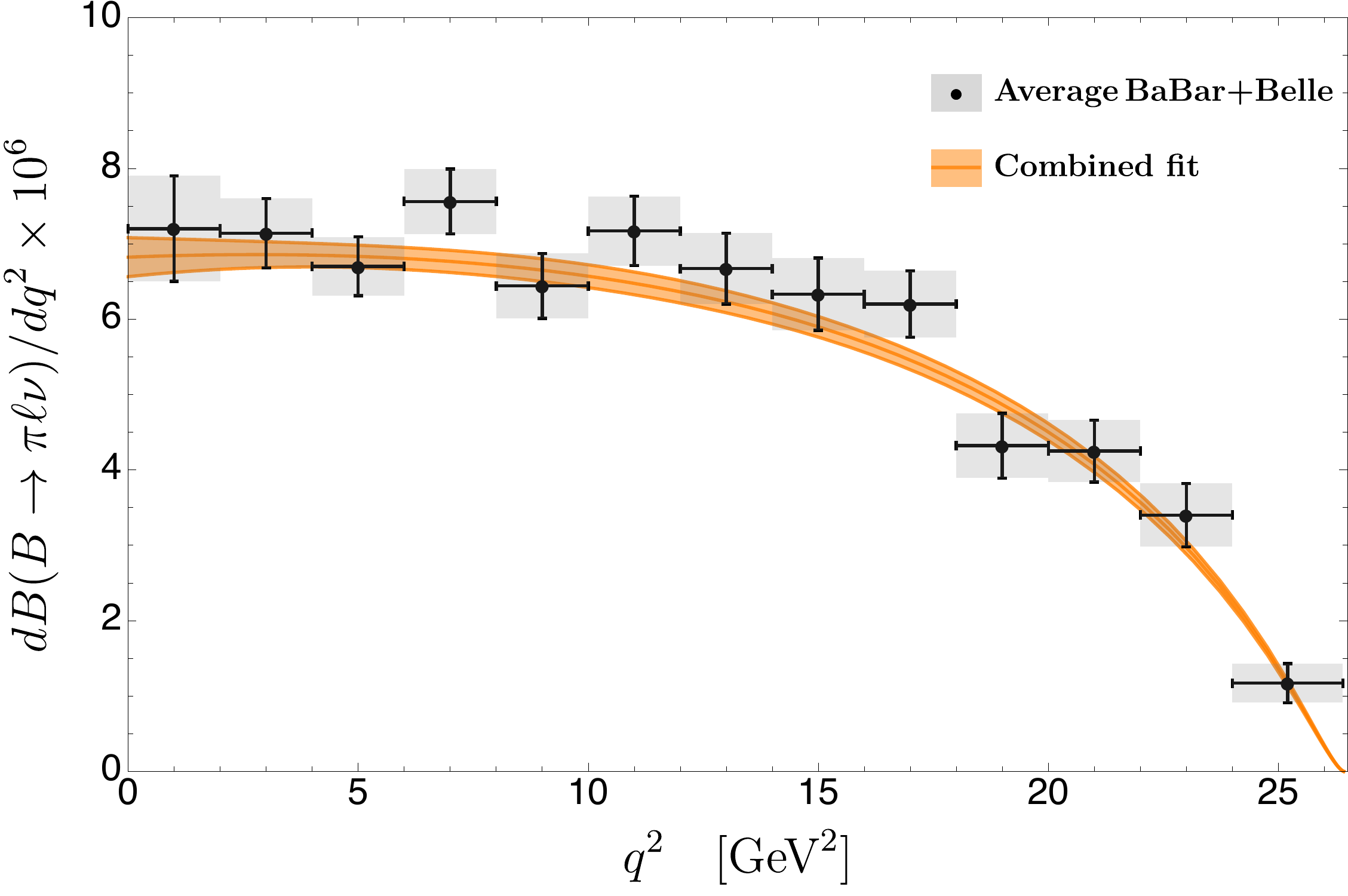}\quad\includegraphics[width=8.855cm]{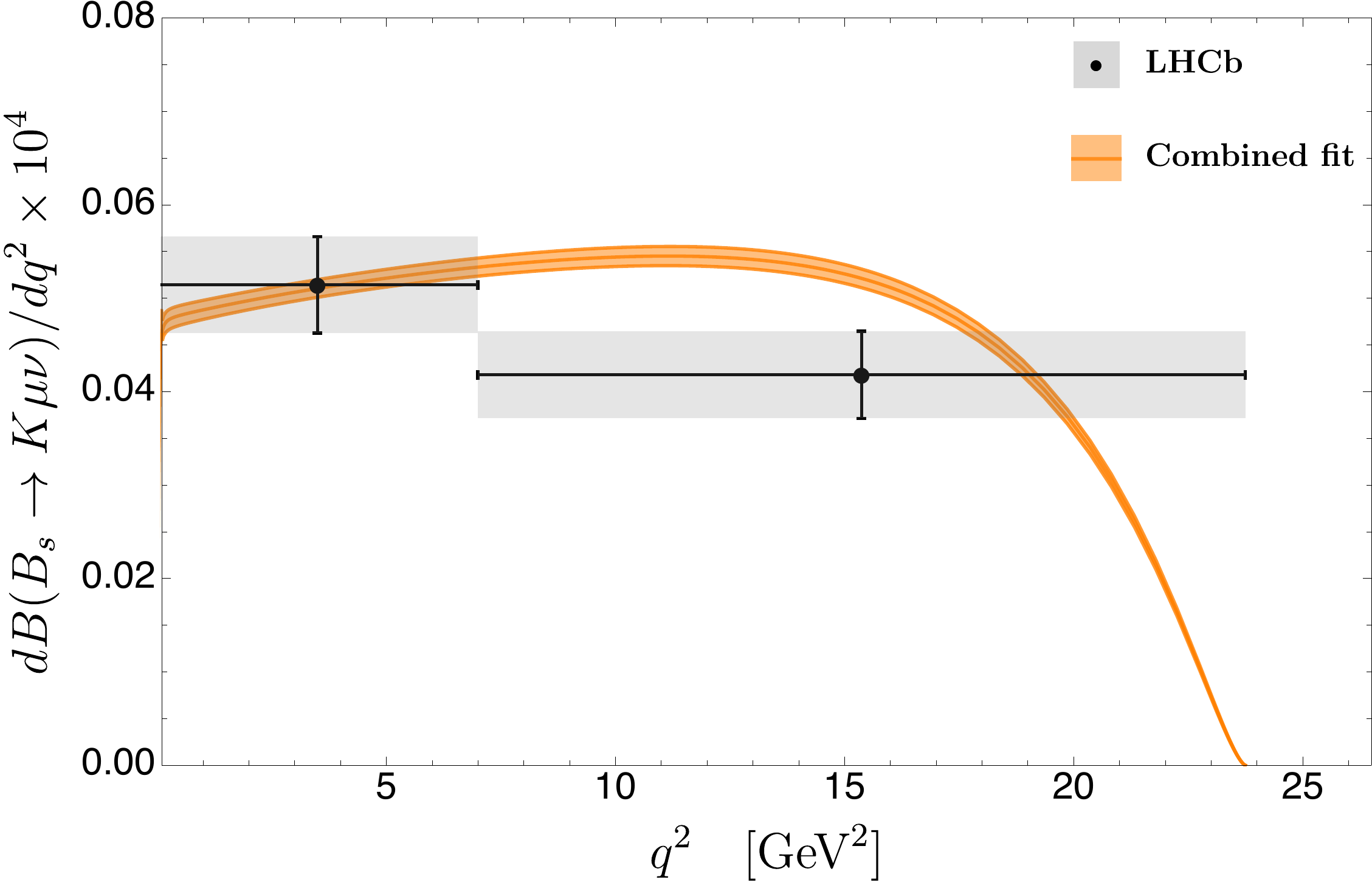} 
\caption{Averaged BaBar and Belle $B\to\pi\ell\nu$ (left) \cite{Amhis:2016xyh} and LHCb $B_{s}\to K^{-}\mu^{+}\nu_{\mu}$ (right) \cite{Aaij:2020nvo} differential branching ratio distributions (gray) as compared to our best fit result (orange) obtained in combined fits to both decays as presented in Table\,\ref{Table:CombinedFit}.
The two LHCb data points are placed in the middle of each bin and have been divided by the bin width.}
\label{Fig:BPiCombined} 
\end{figure*}

\begin{figure*}
\includegraphics[width=8.75cm]{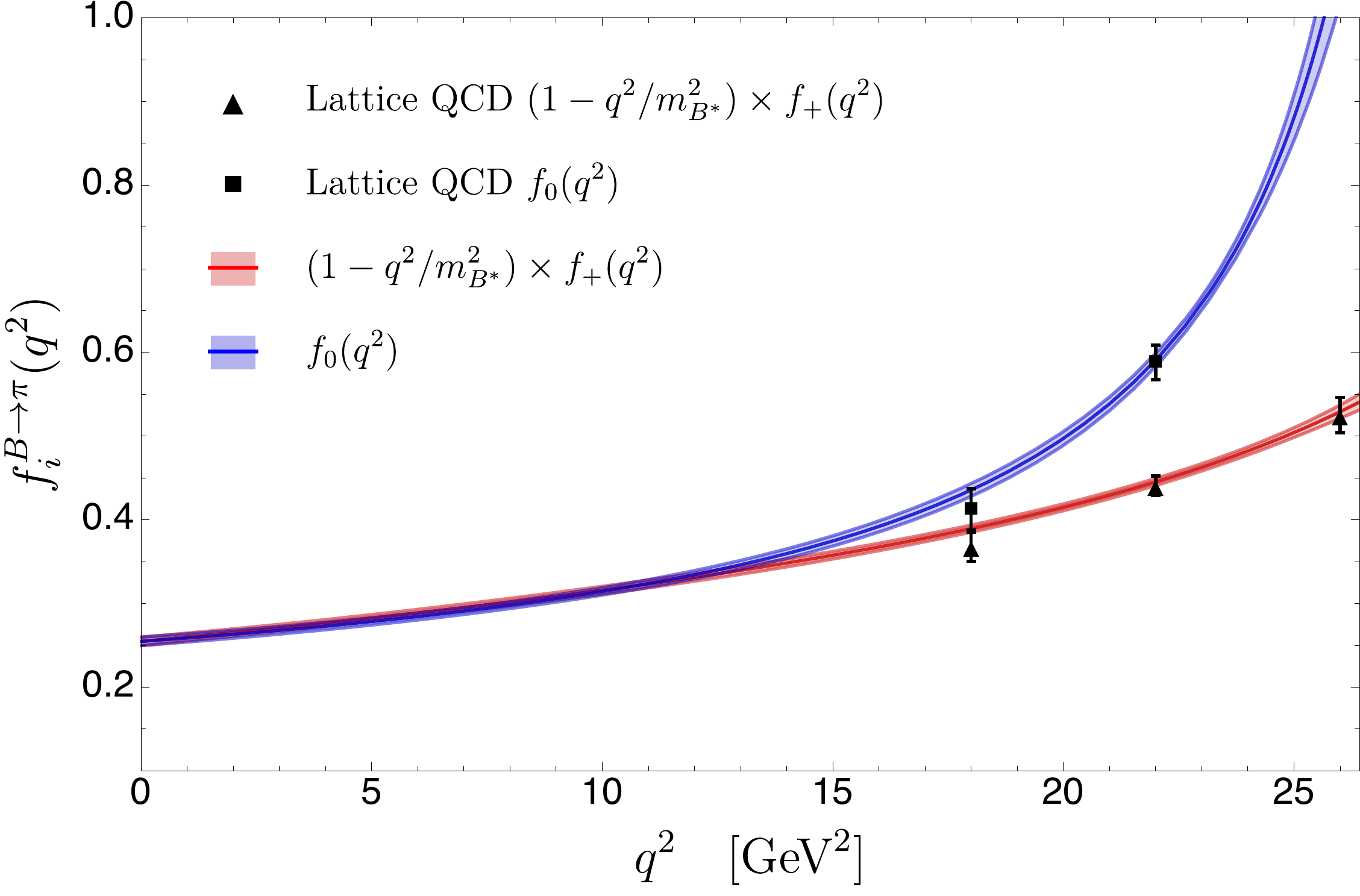}\quad\includegraphics[width=8.75cm]{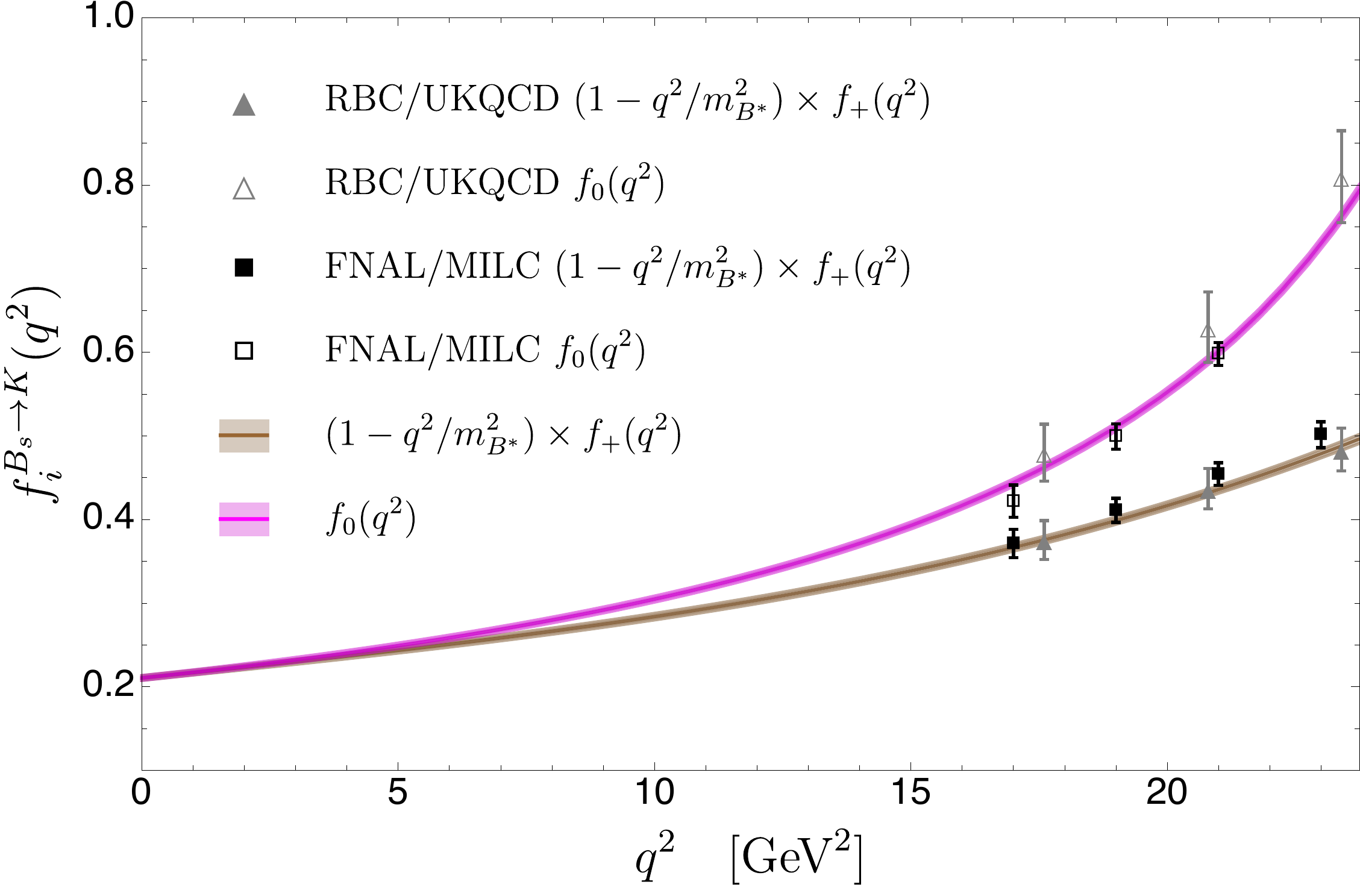}
\caption{Lattice-QCD data for the $B\to\pi$ (left plot) and $B_{s}\to K$ (right plot) vector and scalar form factors compared to our best fit results obtained in combined fits as presented in Table\,\ref{Table:CombinedFit}.}
\label{Fig:FormFactorsCombined} 
\end{figure*}

We would like to note, on the one hand, that our $|V_{ub}|$ result in Eq.\,(\ref{Eq:VubCombined}) corresponds to the most precise determination of $|V_{ub}|$ to date, and that this value is shifted about $1.4\sigma$ downwards with respect to $|V_{ub}|=3.86(11)\times10^{-3}$ extracted from $B\to\pi\ell\nu_{\ell}$ alone (cf.\,Table \ref{Table:FitSM}), and about $1\sigma$ upwards with respect to $|V_{ub}|=3.58(9)\times10^{-3}$ obtained from the individual analysis of the $B_{s}\to K\ell\nu_{\ell}$ channel (cf.\,Table \ref{Table:FitDataAllLatticeBs}).
On the other hand, our determination is far more precise than both the leptonic $B\to\tau\nu_{\tau}$, $|V_{ub}|=4.01(9)(63)\times10^{-3}$ \cite{Aoki:2019cca}, and the inclusive, $|V_{ub}|=4.25(12)^{+15}_{-14}(23)\times10^{-3}$ \cite{Zyla:2020zbs}, determinations, and that 
the tension between our $|V_{ub}|$ result in Eq.\,(\ref{Eq:VubCombined}) and the latter is of about $1.8\sigma$.\footnote{$3.4\sigma$ if the inclusive determination $|V_{ub}|=4.32(12)^{+12}_{-13}\times10^{-3}$ \cite{Amhis:2019ckw} is considered instead, and $1.5\sigma$ with respect to the preliminary value $|V_{ub}|=4.06(9)(16)(15)\times10^{-3}$ in \cite{Cao:2021uwy}.}
The results given in Table \ref{Table:CombinedFit} corresponds, to the best of our knowledge, to the first correlated results between the $B\to\pi$ and $B_{s}\to K$ form factors, which can serve as guidance for those Lattice Collaborations that are planning making available the full theoretical correlation between form factors for different process in their final results \cite{Bazavov:2019aom}.
\begin{table*}
\begin{center}
\begin{tabular}{|l|c|ccccccccccccc|}
\hline
Parameter&Central value&\multicolumn{13}{c|}{\multirow{1}{*}{Correlation matrix}} \\
\hline
$|V_{ub}|\times10^{3}$ &$3.68(5)$&1&$-0.404$&$0.086$&$0.221$&$-0.185$&$0.082$&$-0.082$&$-0.610$&$-0.239$&$0.138$&$-0.150$&$0.203$&$-0.386$\\ 
$a_{0}^{+}$ &$0.255(5)$&&1&$-0.432$&$0.500$&$-0.405$&$-0.745$&$-0.564$&$0.246$&$0.096$&$-0.056$&$0.061$&$-0.082$&$0.156$\\ 
$a_{1}^{+}\times10^{3}$ &$-1.36(60)$&&&$1$&$0.055$&$-0.331$&$0.186$&$0.048$&$-0.053$&$-0.021$&$0.012$&$-0.013$&$0.018$&$-0.033$\\ 
$a_{2}^{+}\times10^{4}$ &$-0.66(68)$&&&&$1$&$-0.957$&$-0.750$&$-0.821$&$-0.135$&$-0.053$&$0.031$&$-0.033$&$0.045$&$-0.085$\\ 
$m_{B^{*}(1^{-})}$ pole(s) [GeV]  &$=5.325\&6.24$&&&&&$1$&$0.685$&$0.775$&$0.113$&$0.044$&$-0.026$&$0.028$&$-0.038$&$0.071$\\ 
$a_{1}^{0}\times10^{2}$ &$-0.46(6)$&&&&&&$1$&$0.962$&$-0.050$&$-0.020$&$0.011$&$-0.012$&$0.017$&$-0.032$\\ 
$m_{B^{*}(0^{+})}$ pole(s) [GeV]  &$5.45$&&&&&&&$1$&$0.050$&$0.020$&$-0.011$&$0.012$&$-0.017$&$0.032$\\ 
\hline
$c_{0}^{+}$ &$0.211(3)$&&&&&&&&1&$-0.052$&$0.095$&$-0.046$&$0.030$&$0.765$\\ 
$c_{1}^{+}\times10^{3}$ &$4.96(2.32)$&&&&&&&&&$1$&$-0.975$&$0.968$&$-0.992$&$-0.121$\\ 
$c_{2}^{+}\times10^{4}$ &$-0.37(26)$&&&&&&&&&&$1$&$-0.994$&$0.989$&$0.185$\\ 
$c_{3}^{+}\times10^{5}$ &$0.81(43)$&&&&&&&&&&&$1$&$-0.990$&$-0.115$\\ 
$m_{B^{*}(1^{-})}$ pole(s) [GeV]  &$=5.325\&12.13$&&&&&&&&&&&&$1$&$0.088$\\ 
$m_{B^{*}(0^{+})}$ pole(s) [GeV]  &$5.69$&&&&&&&&&&&&&$1$\\ 
\hline
\end{tabular}
\caption{Best fit values, uncertainties and correlation matrix for the output quantities of our $\chi^{2}$ fits Eq.\,(\ref{Eq:chi2Combined}) obtained from a combined fit to the averaged $B\to\pi\ell\nu_{\ell}$ BaBar and Belle \cite{Amhis:2016xyh} and the $B_{s}\to K^{-}\ell^{+}\nu_{\ell}$ LHCb \cite{Aaij:2020nvo} experimental data in combination with the Lattice-QCD $B\to\pi$ \cite{Aoki:2019cca} and $B_{s}\to K$ \cite{Flynn:2015mha,Bazavov:2019aom} vector and scalar form factors simulations.}
\label{Table:CombinedFit}
\end{center}
\end{table*}
The results of the combined fit are plotted in Fig.\,\ref{Fig:BPiCombined} for the differential $B\to\pi\ell\nu_{\ell}$ (left plot) and $B_{s}\to K^{-}\mu^{+}\nu_{\mu}$ (right plot) branching ratio distributions, and in Fig.\,\ref{Fig:FormFactorsCombined} for the corresponding vector and scalar form factors. 
Concerning the form factor values at $q^{2}=0$, we obtain:
\begin{eqnarray}
f_{+,0}^{B\pi}(0)=0.255(5)\,,\quad f_{+,0}^{B_{s}K}(0)=0.211(3)\,,
\end{eqnarray}
which can be compared with the output values: $f_{+,0}^{B\pi}(0)=0.253(11)$ \cite{Lattice:2015tia} and $f_{+,0}^{B_{s}K}(0)=0.135(50)$ \cite{Bazavov:2019aom} from the FNAL/MILC Lattice Collaborations; $f_{+,0}^{B\pi}(0)=0.26^{+0.04}_{-0.03}$ \cite{Duplancic:2008ix} and $f_{+,0}^{B_{s}K}(0)=0.30^{+0.04}_{-0.03}$ \cite{Duplancic:2008tk}, $f_{+,0}^{B\pi}(0)=0.301(23)$ and $f_{+,0}^{B_{s}K}(0)=0.336(23)$ \cite{Khodjamirian:2017fxg}, and $f_{+}^{B\pi}(0)=0.252^{+0.019}_{-0.028}$ \cite{Bharucha:2012wy} from light-cone sum rules; $f_{+,0}^{B\pi}(0)=f_{+,0}^{B_{s}K}(0)=0.26^{+0.04}_{-0.03}\pm0.02$ from perturbative QCD \cite{Wang:2012ab}; and $f_{+,0}^{B_{s}K}(0)=0.284(14)$ from relativistic quark model \cite{Faustov:2013ima}.

\begin{figure}
\includegraphics[width=8.75cm]{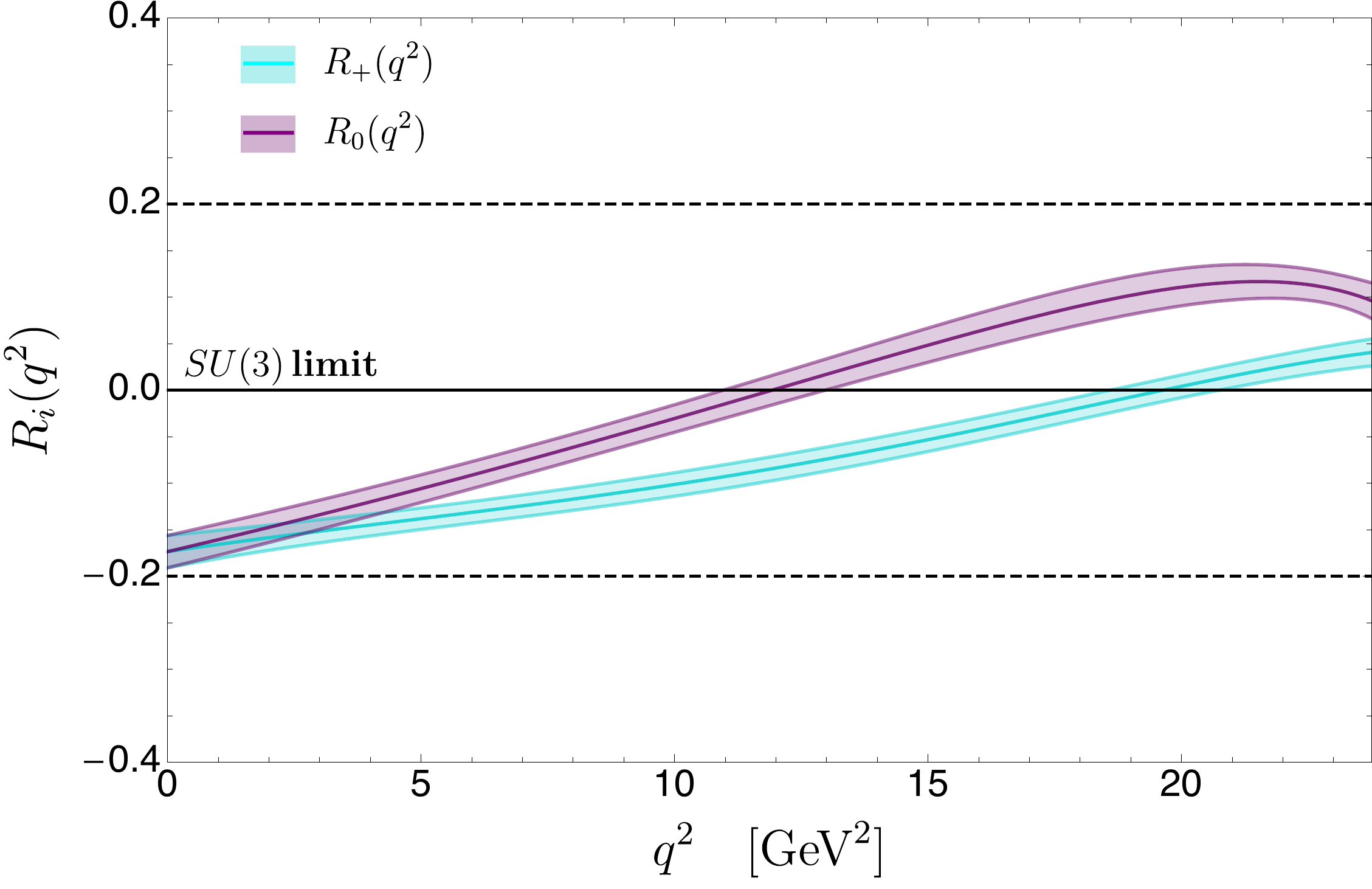}
\caption{$SU(3)$-breaking ratios $R_{+}(q^{2})$ (cyan) and $R_{0}(q^{2})$ (purple) (cf.\,Eq.\,(\ref{Eq:SU(3)})) using our determinations of the $B\to\pi$ and $B_{s}\to K$ vector and scalar form factors from Table \ref{Table:CombinedFit}.}
\label{Fig:SU3breaking} 
\end{figure}

Finally, in Fig.\,\ref{Fig:SU3breaking} we present results for the quantity:
\begin{equation}
R_{i}(q^{2})=\frac{f_{i}^{B_{s}K}(q^{2})}{f_{i}^{B\pi}(q^{2})}-1\,,
\label{Eq:SU(3)}
\end{equation} 

with $i=+,0$, which provides a measure of $SU(3)$-breaking.\footnote{In the $SU(3)$ limit, {\it{i.e.}} $m_{d}=m_{s}$, the $B\to\pi$ and $B_{s}\to K$ form factors should be identical.}
As seen, while the results for $R_{+}(q^{2})$ (cyan) and $R_{0}(q^{2})$ (purple) are similar at low energies $(q^{2}\lesssim5$ GeV$^{2}$), $R_{0}(q^{2})$ is larger than $R_{+}(q^{2})$ at higher energies, and the deviations from unity are consistent with the simple counting $(m_{s}-m_{d})/\Lambda_{\rm{QCD}}\sim20\%$.

%% file: Section4Pheno.tex
\section{Phenomenological applications}\label{section4}

As a benefit of our results of Table \ref{Table:CombinedFit}, we provide calculations for different phenomenological observables such as total decay rates, ratio of $\tau$-to-$\mu$ differential decay rates or the forward-backward asymmetry, and its normalized version.

Integrating the differential decay rates (cf.\,Eq.\,(\ref{Eq:DecayDistribution})) over the kinematically-allowed $q^{2}$ ranges, and dividing by $|V_{ub}|^{2}$, we obtain: 
\begin{eqnarray}
\Gamma(B\to\pi\mu\nu_{\mu})/|V_{ub}|^{2}&=&6.90(16)\,\,{\rm{ps}}^{-1}\,,\\[1ex]
\Gamma(B\to\pi\tau\nu_{\tau})/|V_{ub}|^{2}&=&4.55(9)\,\,{\rm{ps}}^{-1}\,,\\[1ex]
\Gamma(B_{s}\to K\mu\nu_{\mu})/|V_{ub}|^{2}&=&5.31(13)\,\,{\rm{ps}}^{-1}\,,\\[1ex]
\Gamma(B_{s}\to K\tau\nu_{\tau})/|V_{ub}|^{2}&=&3.70(8)\,\,{\rm{ps}}^{-1}\,,
\end{eqnarray}
with errors of only about $2\%$.

\begin{figure}
\includegraphics[width=8.75cm]{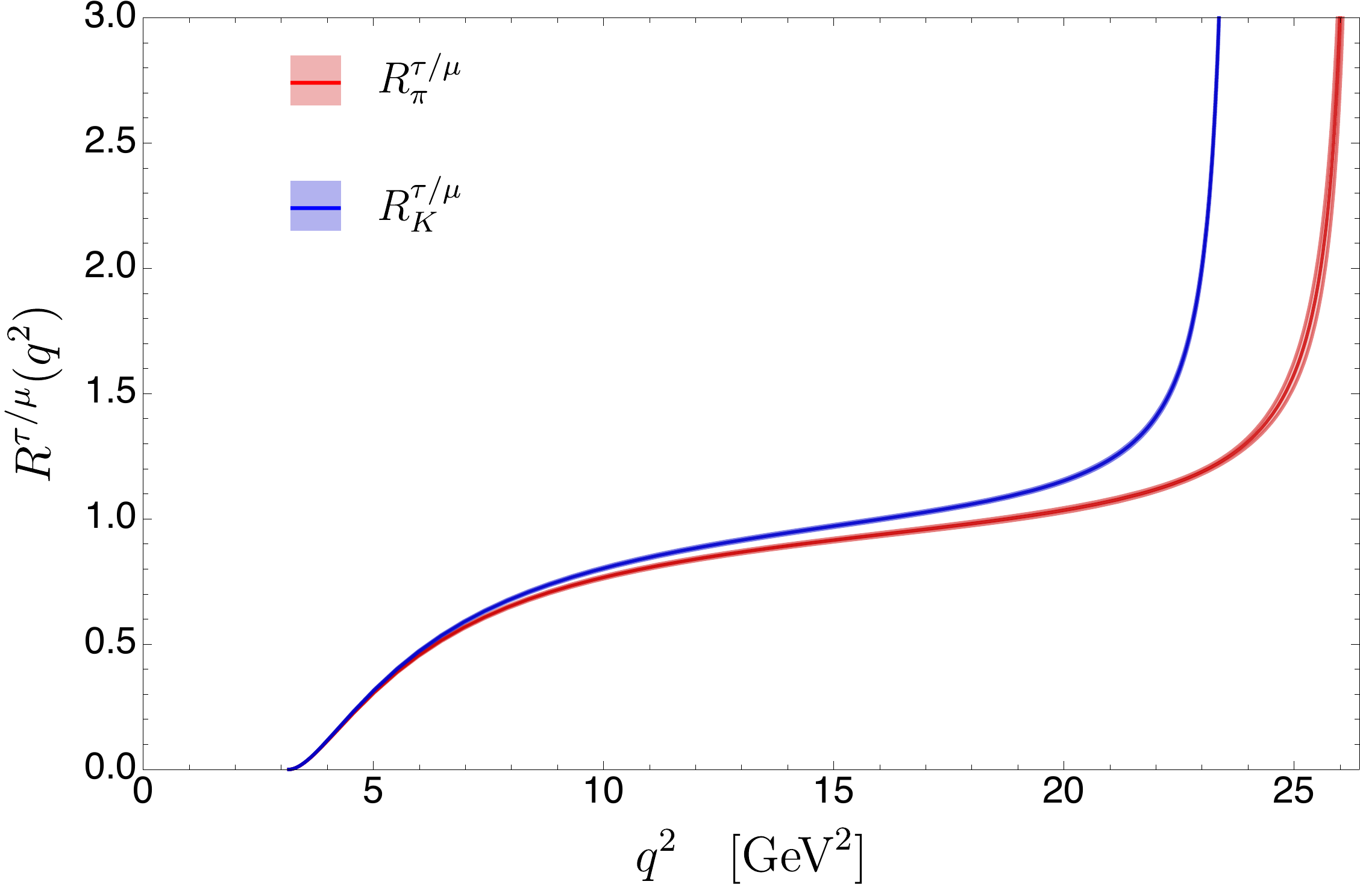}
\caption{Standard Model ratio of differential $\tau$-to-$\mu$ decay rates Eq.\,(\ref{Eq:R}) using our determinations of the $B\to\pi$ and $B_{s}\to K$ vector and scalar form factors from Table \ref{Table:CombinedFit}.}
\label{Fig:R} 
\end{figure}

The $\tau$-to-$\mu$ $q^{2}$-dependent ratio of differential decay rates
\begin{equation}
\mathcal{R}_{\pi(K)}^{\tau/\mu}(q^{2})=\frac{d\Gamma(B_{(s)}\to\pi(K)\tau\nu_{\tau})/dq^{2}}{d\Gamma(B_{(s)}\to\pi(K)\mu\nu_{\mu})/dq^{2}}\,,
\label{Eq:R}
\end{equation}
and its integrated form
\begin{equation}
R_{\pi(K)}^{\tau/\mu}=\frac{\int_{m_{\tau}^{2}}^{(m_{B_{(s)}}-m_{\pi(K)})^{2}}dq^{2}d\Gamma(B_{(s)}\to\pi(K)\tau\nu_{\tau})/dq^{2}}{\int_{m_{\mu}^{2}}^{(m_{B_{(s)}}-m_{\pi(K)})^{2}}dq^{2}d\Gamma(B_{(s)}\to\pi(K)\mu\nu_{\mu})/dq^{2}}\,,
\label{Eq:Rint}
\end{equation}
can be used for precise Standard Model test that is independent of $|V_{ub}|$.
Fig.\,\ref{Fig:R} shows our predictions for Eq.\,(\ref{Eq:R}) using our $B\to\pi\ell\nu_{\ell}$ and $B_{s}\to K\mu\nu_{\mu}$ form factor outputs from Table \ref{Table:CombinedFit}, while our numerical predictions for Eq.\,(\ref{Eq:Rint}) are found to be:
\begin{eqnarray}
R_{\pi}^{\tau/\mu}&=0.660(5)\,,\\[1ex]
R_{K}^{\tau/\mu}&=0.697(3)\,,
\label{Eq:Rvalues}
\end{eqnarray}
which are only $1\%$ error.
These values are found to be in agreement with, but more precise than, $R_{\pi}^{\tau/\mu}=0.69(19)$ and $R_{K}^{\tau/\mu}=0.77(12)$ from Ref.\,\cite{Flynn:2015mha}, and $R_{K}^{\tau/\mu}=0.77(6)$ from Ref.\,\cite{Bazavov:2019aom}\footnote{In \cite{Bazavov:2019aom}, the value $R_{K}^{\tau/\mu}=0.836(34)$ is reported, which corresponds to taking $m_{\tau}^{2}$ as the lower limit of integration in the denominator of Eq.\,(\ref{Eq:Rint}).}. 

\begin{figure*}
\includegraphics[width=18cm]{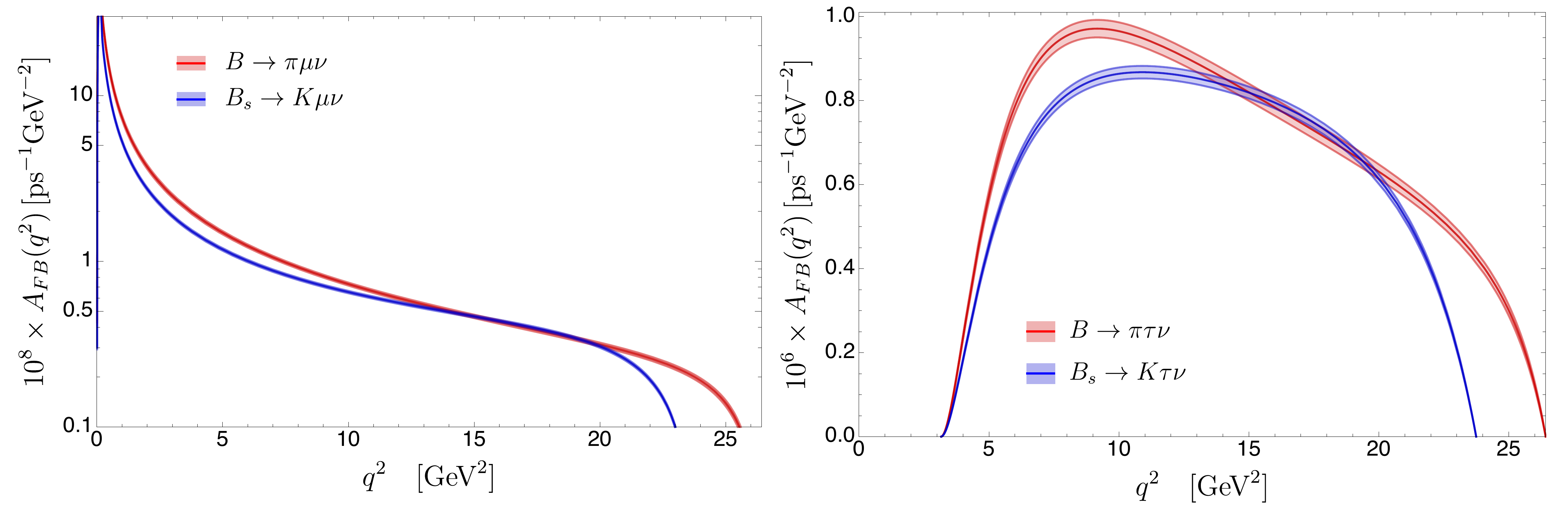}
\caption{Predictions for the forward-backward asymmetry Eq.\,(\ref{Eq:AFB}) for $B\to\pi\mu\nu$ and $B_{s}\to K\mu\nu$ (left), and $B\to\pi\tau\nu$ and $B_{s}\to K\tau\nu$ (right), using our fit results from Table \ref{Table:CombinedFit}.}
\label{Fig:AFB} 
\end{figure*}

\begin{figure*}
\includegraphics[width=18cm]{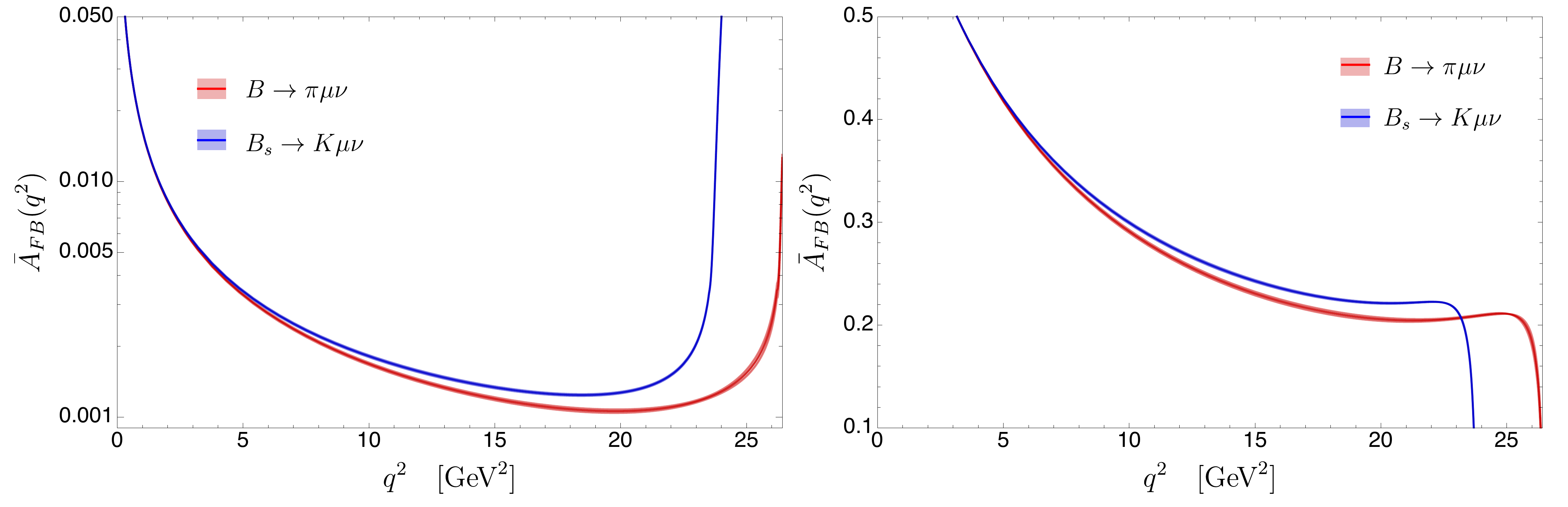}
\caption{Predictions for the normalized forward-backward asymmetry Eq.\,(\ref{Eq:AFBnorm}) for $B\to\pi\mu\nu$ and $B_{s}\to K\mu\nu$ (left), and $B\to\pi\tau\nu$ and $B_{s}\to K\tau\nu$ (right), using our fit results from Table \ref{Table:CombinedFit}.}
\label{Fig:AFBNorm} 
\end{figure*}

Concerning the forward-backward asymmetry, $A_{FB}$, it is a quantity sensitive to the mass of the final-state charged lepton and its theoretical expression is given by:
\begin{eqnarray}
A_{FB}^{B_{(s)}\to\pi(K)\ell\nu_{\ell}}(q^{2})&\equiv&\left(\int_{0}^{1}-\int_{-1}^{0}\right)d\cos\theta_{\ell}\frac{d^{2}\Gamma(B_{(s)}\to\pi(K)\ell\nu_{\ell})}{dq^{2}d\cos\theta_{\ell}}\nonumber\\[1ex]
&=&\frac{G_{F}^{2}|V_{ub}|^{2}}{32\pi^{3}m_{B_{(s)}}}\left(1-\frac{m_{\ell}^{2}}{q^{2}}\right)^{2}|\vec{p}_{\pi(K)}|^{2}\nonumber\\[1ex]
&\times&\frac{m_{\ell}^{2}}{q^{2}}(m_{B_{(s)}}^{2}-m_{\pi(K)}^{2}){\rm{Re}}[f_{+}(q^{2})f_{0})(q^{2})]\,,
\label{Eq:AFB}
\end{eqnarray}
where $\theta_{\ell}$ is the angle between the charged-lepton and the $B_{(s)}$-meson momenta in the $q^{2}$ rest frame.
In Fig.\ref{Fig:AFB}, we show our predictions for $A_{FB}$ using our best fit results from \ref{Table:CombinedFit}.
Integrating over the corresponding kinematic $q^{2}$ ranges, and diving by $|V_{ub}|^{2}$, we obtain:
\begin{eqnarray}
&&\int_{m_{\mu}^{2}}^{(m_{B}-m_{\pi})^{2}}dq^{2}A_{FB}^{B\to\pi\mu\nu}(q^{2})/|V_{ub}|^{2}=0.034(1)\,\,{\rm{ps}}^{-1}\,,\\
&&\int_{m_{\tau}^{2}}^{(m_{B}-m_{\pi})^{2}}dq^{2}A_{FB}^{B\to\pi\tau\nu}(q^{2})/|V_{ub}|^{2}=1.16(3)\,\,{\rm{ps}}^{-1}\,,\\
&&\int_{m_{\mu}^{2}}^{(m_{B_{s}}-m_{K})^{2}}dq^{2}A_{FB}^{B_{s}\to K\mu\nu}(q^{2})/|V_{ub}|^{2}=0.0255(6)\,\,{\rm{ps}}^{-1}\,,\nonumber\\
&&\\
&&\int_{m_{\tau}^{2}}^{(m_{B_{s}}-m_{K})^{2}}dq^{2}A_{FB}^{B_{s}\to K\tau\nu}(q^{2})/|V_{ub}|^{2}=0.99(2)\,\,{\rm{ps}}^{-1}\,,
\end{eqnarray}
with errors of about $3\%$.
While these values are in general agreement with, but more precise than, those in Ref.\,\cite{Flynn:2015mha}, our results show a difference of about $1.5\sigma$ with \cite{Bazavov:2019aom}.
Finally, the normalized forward-backward asymmetry,
\begin{eqnarray}
\bar{A}_{FB}^{B_{(s)}\to\pi(K)\ell\nu_{\ell}}(q^{2})\equiv\frac{\int_{m_{\ell}^{2}}^{(m_{B(s)}^{2}-m_{\pi(K)}^{2})^{2}}dq^{2}A_{FB}^{B_{(s)}\to\pi(K)\ell\nu_{\ell}}(q^{2})}{\int_{m_{\ell}^{2}}^{(m_{B(s)}^{2}-m_{\pi(K)}^{2})^{2}}dq^{2}d\Gamma(B_{(s)}\to\pi(K)\ell\nu_{\ell})/dq^{2}}\,,\nonumber\\
\label{Eq:AFBnorm}
\end{eqnarray}
is an interesting observable as it is independent of $|V_{ub}|$.
Our predictions are show in Fig.\,\ref{Fig:AFBNorm}, whereas integrating Eq.\,(\ref{Eq:AFBnorm}) over the allowed $q^{2}$ ranges we find:
\begin{eqnarray}
&&\bar{A}_{FB}^{B\to\pi\mu\nu}=0.0049(1)\,,\\
&&\bar{A}_{FB}^{B\to\pi\tau\nu}=0.255(1)\,,\\
&&\bar{A}_{FB}^{B_{s}\to K\mu\nu}=0.0048(1)\,,\\
&&\bar{A}_{FB}^{B_{s}\to K\tau\nu}=0.2684(9)\,,
\end{eqnarray}
with errors of about $2\%$ and $1\%$ for $\mu$ and $\tau$, respectively.
While these values are found to be in agreement with Ref.\,\cite{Flynn:2015mha}, our results are more precise. 
With respect to \cite{Bazavov:2019aom}, our results differ by about $\sim1.6-2.1\sigma$ for $\bar{A}_{FB}^{B_{s}\to K\mu\nu}$ and $\bar{A}_{FB}^{B_{s}\to K\tau\nu}$, respectively.

%% file: SectionLastOutlook.tex
\section{Outlook}\label{conclusions}

In this work we have explored the role of the decay $B_{s}\to K\ell\nu_{\ell}$ in complementing the traditional channel $B\to\pi\ell\nu_{\ell}$ in the determination of the CKM element $|V_{ub}|$.
The motivation of this study is the first reported measurement of the branching ratio of the decay $B_{s}\to K^{-}\mu^{+}\nu_{\mu}$ by the LHCb Collaboration \cite{Aaij:2020nvo}, making this analysis of timely interest.

Our analysis has been based on the method of Pad\'{e} approximants to the corresponding form factors, and proceeded in three steps.
First, we used the most precise measurements of the differential $B\to\pi\ell\nu_{\ell}$ branching ratio distribution given by BaBar and Belle, along with the Lattice-QCD calculations of the $B\to\pi$ vector and scalar form factors, to extract $|V_{ub}|$ from a combined fit which makes use of both information sets in a democratic way.
As a result of this exercise we have obtained $|V_{ub}|=3.86(11)\times10^{-3}$ (cf.\,Eq.\,(\ref{Eq:BestFitBtoPi})), together with the form factor parameters and their correlation matrix collected in Table \ref{Table:CorrelationBestFitBtoPi} of Appendix \ref{Appendix1}.
We note that our result for $|V_{ub}|$ differs only by about $1.35\sigma$ with the determination from inclusive decays $B\to X_{u}\ell\nu_{\ell}$, $|V_{ub}|=4.25(12)^{+15}_{-14}(23)\times10^{-3}$ \cite{Zyla:2020zbs},\footnote{$1.2\sigma$ with respect to the preliminary value $|V_{ub}|=4.06(9)(16)(15)\times10^{-3}$ \cite{Cao:2021uwy}.} confirming the trend of obtaining higher values of $|V_{ub}|$ from recent exclusive $B\to\pi\ell\nu_{\ell}$ determinations \cite{Biswas:2021qyq,Leljak:2021vte}.
Second, we have determined $|V_{ub}|$ from the decay $B_{s}\to K\ell\nu_{\ell}$ performing combined fits to the experimental LHCb data and Lattice input on the $B_{s}\to K$ form factors.
Our fits yield $|V_{ub}|=3.58(9)\times10^{-3}$ and the form factor parameters and their correlation matrix given in Table \ref{Table:FitDataAllLatticeBs}.
This is a relevant result, as the central $|V_{ub}|$ value from $B_{s}\to K\ell\nu_{\ell}$ suffers a shift of about $1.9\sigma$ downwards with respect to the one obtained from $B\to\pi\ell\nu_{\ell}$, thus increasing the difference with respect to the determination from inclusive decays to $2.1\sigma$. We traced back this difference to the impact of existing experimental data used in each channel: Lattice input in form factors in both channels tend to yield values for $|V_{ub}|$ around $3.6\times 10^{-3}$ while experimental data seem to prefer higher values of around $|V_{ub}|=3.9(9)\times10^{-3}$. Since experimental data for the $B_{s}\to K$ is scarce, that channel is dominated by Lattice input thus confronting the $B \to \pi$ one.
Third, and last, we have performed a simultaneous analysis to all available experimental and Lattice-QCD information on both $B\to\pi\ell\nu_{\ell}$ and $B_{s}\to K^{-}\mu^{+}\nu_{\mu}$ decays.
The resulting fit yields $|V_{ub}|=3.68(5)\times10^{-3}$, which is a $1.4\%$ error and differs by only $1.8\sigma$ from the inclusive value.

\begin{figure*}
\includegraphics[width=8.75cm]{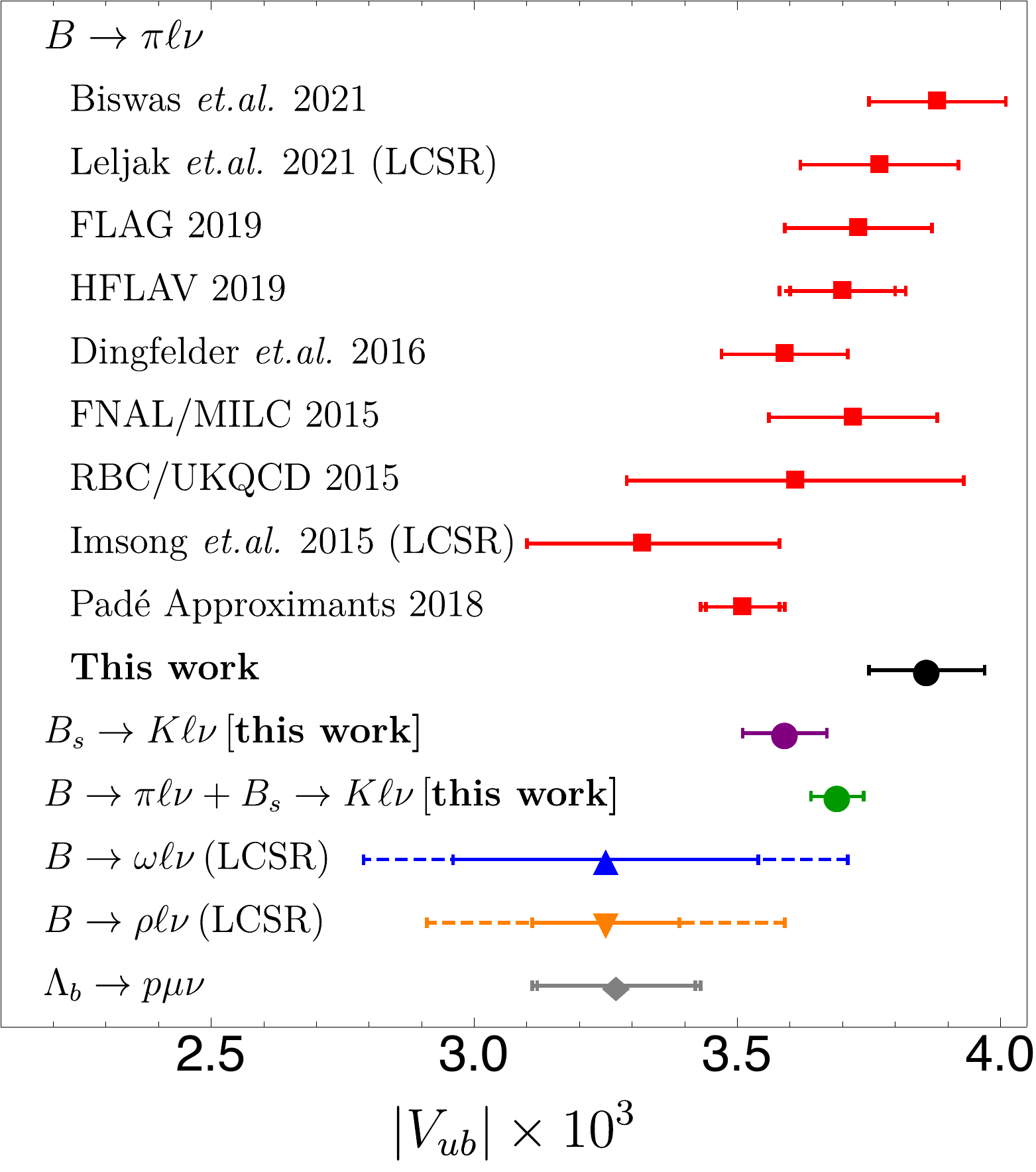}
\caption{Status of $|V_{ub}|$ determinations from exclusive $B\to\pi\ell\nu_{\ell}$ decays (red squares) including Biswas et.\,al. \cite{Biswas:2021qyq}, Leljak et.\,al. \cite{Leljak:2021vte}, FLAG 2019 \cite{Aoki:2019cca}, HFLAV 2019 \cite{Amhis:2019ckw}, Dingfelder et.\,al. \cite{Dingfelder:2016twb}, FNAL/MILC 2015 \cite{Lattice:2015tia}, RBC/UKQCD \cite{Flynn:2015mha}, Imsong et.\,al. \cite{Imsong:2014oqa}, Pad\'{e} approximants \cite{Gonzalez-Solis:2018ooo} and this work (black circle), from $B_{s}\to K\ell\nu_{\ell}$ (this work, purple circle), from a combination of $B\to\pi\ell\nu_{\ell}$ and $B_{s}\to K\ell\nu_{\ell}$ decays (this work, green circle), from $B\to\omega\ell\nu_{\ell}$ (upward blue triangle) and $B\to\rho\ell\nu_{\ell}$ (downward orange triangle) \cite{Straub:2015ica}, and from $\Lambda_{b}\to p\mu\nu_{\mu}$ (gray diamond) LHCb \cite{Aaij:2015bfa}.}
\label{Fig:VubStatus} 
\end{figure*}

The process of performing a combined fit to both decays also tests for their compatibility, and the result is a $|V_{ub}|$ that stays $\sim1\sigma$ away from the $|V_{ub}|$ results extracted from the individual decay modes.
In this sense, more precise measurements of the differential $B_{s}\to K\ell\nu_{\ell}$ decay distribution with a finer resolution of the $q^{2}$ bins will help achieve more conclusive results.
Our value is presented and compared with other determinations using different methods and fitted data sets in Fig.\,\ref{Fig:VubStatus}.
As seen, our value is the most precise to date.
The coefficients of the Pad\'{e} approximants for the $B\to\pi$ and $B_{s}\to K$ form factor are given in Table \ref{Table:CombinedFit} together with their correlation matrix.
The latter represents, to the best of our knowledge, the first correlated results for these form factors.
As a benefit of our analysis, in Sec. \ref{section4} we have calculated different phenomenological observables such as total decay rates, ratio of $\tau$-to-$\mu$ differential decay rates or the forward-backward asymmetry, and its normalized version, with an accuracy of few $\%$.

On the experimental side, the decay $B_{s}\to K\ell\nu_{\ell}$ is expected to be studied at the Belle-II experiment \cite{Kou:2018nap}, which due to the $e^{+}e^{-}$ collisions offers a cleaner environment than the LHCb. 
Belle II will collect a large numbers of $B_{s}$-meson pairs and, although measurements of the $B_{s}\to K\ell\nu_{\ell}$ rates are not expected to reach the experimental accuracy of their results for $B\to\pi\ell\nu_{\ell}$ \cite{Abudinen:2020zfj}, more precise measurements will clearly help improve the determination of $|V_{ub}|$.
On the Lattice front, there are plans to reduce the contributions from the dominant sources of statistical \cite{Flynn:2015mha} and systematic \cite{Gelzer:2019zwx,Bazavov:2019aom} uncertainties in upcoming form factor calculations, as well as making the full correlation matrix between the $B\to\pi$ and $B_{s}\to K$ form factors available \cite{Bazavov:2019aom,Flynn:2020nmk}.
These, and other improvements, will allow to obtain form factors with percent level precision and hence allow for exclusive $|V_{ub}|$ determinations with improved precision.

%% file: Appendix1.tex
\section{Fit results and form factors simulations}\label{Appendix1}

\begin{table*}
\begin{center}
\begin{tabular}{|l|c|ccccccc|}
\hline
Parameter&Central value&\multicolumn{7}{c|}{\multirow{1}{*}{Correlation matrix}} \\
\hline
$|V_{ub}|\times10^{3}$ &$3.86(11)$&1&$-0.571$&$0.035$&$0.324$&$-0.251$&$0.139$&$-0.121$\\ 
$a_{0}^{+}$ &$0.247(8)$&&1&$-0.374$&$0.341$&$-0.298$&$-0.719$&$-0.473$\\ 
$a_{1}^{+}\times10^{3}$ &$-1.3(8)$&&&$1$&$0.028$&$-0.297$&$0.220$&$0.105$\\ 
$a_{2}^{+}\times10^{4}$ &$-0.3(1.0)$&&&&$1$&$-0.958$&$-0.681$&$-0.795$\\ 
$m_{B^{*}(1^{-})}$ pole(s) [GeV]  &$=5.325\&6.46$&&&&&$1$&0.633&$0.747$\\ 
$a_{1}^{0}\times10^{2}$ &$-0.4(1)$&&&&&&$1$&0.943\\ 
$m_{B^{*}(0^{+})}$ pole(s) [GeV]  &$5.44$&&&&&&&$1$\\ 
\hline
\end{tabular}
\caption{Best fit values, uncertainties and correlation matrix for the output quantities of our best $\chi^{2}_{B\pi}$ fit Eq.\,(\ref{Eq:chi2}) obtained with the Pad\'{e} element $P_{1,1}^{2}$ (cf.\,Table\,\ref{Table:FitSMPadeType}).}
\label{Table:CorrelationBestFitBtoPi}
\end{center}
\end{table*}

\begin{table}
\begin{center}
\begin{tabular}{|cc|c|ccc|ccc|}
\hline
&&&\multicolumn{3}{c|}{\multirow{1}{*}{Correlation matrix}}\\
Form factor&&&\multicolumn{3}{c|}{\multirow{1}{*}{$f_{+}^{B\pi}$}}  \\
&$q^{2}$ [GeV$^{2}$]&Central values&18&22&26\\
\hline
\multirow{3}{*}{$f_{+}^{B\pi}$}&18&1.102(44)&1&0.757&0.563\\
&22&1.964(54)&&1&0.400\\
&26&5.848(226)&&&1\\
\hline
\end{tabular}
\caption{Central values, errors and correlation matrix for the $B\to\pi$ vector form factor, $f_{+}(q^{2})$, generated at three representative values of $q^{2}$ from the FLAG \cite{Aoki:2019cca} results and used in our fits in Table\,\ref{Table:FitSMvectorSequence}.}
\label{Table:CoefficientsVectorAlone}
\end{center}
\end{table}

\begin{table*}
\begin{center}
\begin{tabular}{|l|c|ccccccc|}
\hline
Parameter&Central value&\multicolumn{7}{c|}{\multirow{1}{*}{Correlation matrix}} \\
\hline
$|V_{ub}|\times10^{3}$ &$3.58(9)$&1&$-0.674$&$-0.332$&$0.168$&$-0.254$&$0.306$&$-0.445$\\ 
$a_{0}^{+}$ &$0.214(5)$&&1&$0.035$&$0.056$&$0.004$&$-0.028$&$0.784$\\ 
$a_{1}^{+}\times10^{3}$ &$6.70(5.40)$&&&$1$&$-0.945$&$0.982$&$-0.997$&$-0.061$\\ 
$a_{2}^{+}\times10^{4}$ &$-0.48(46)$&&&&$1$&$-0.988$&$0.963$&$0.149$\\ 
$a_{3}^{+}\times10^{5}$ &$1.04(96)$&&&&&$1$&$-0.992$&$-0.077$\\ 
$m_{B^{*}(1^{-})}$ pole(s) [GeV]  &$=5.325\&29.5$&&&&&&$1$&0.037\\ 
$m_{B^{*}(0^{+})}$ pole(s) [GeV]  &$5.70$&&&&&&&$1$\\ 
\hline
\end{tabular}
\caption{Best fit values, uncertainties and correlation matrix for the output quantities of our best $\chi^{2}_{B_{s}K}$ fit Eq.\,(\ref{Eq:chi2Bs}) obtained with the Pad\'{e} element $P_{1,1}^{3}$ (cf.\,Table\,\ref{Table:FitDataAllLatticeBs}).}
\label{Table:CorrelationBestFitBstoK}
\end{center}
\end{table*}